\documentclass[journal, onecolumn]{IEEEtran}
\usepackage{graphicx}
\usepackage{amsmath,amsfonts}
\usepackage{multicol}
\usepackage{mathrsfs}
\usepackage{mathcomp}
\usepackage{cite}
\usepackage{array}
\usepackage{multirow}
\usepackage{xcolor}
\usepackage{caption}
\usepackage{subcaption} 
\usepackage{calc}
\usepackage{verbatim}
\usepackage[utf8]{inputenc}
\usepackage{float}
\usepackage[font={small}]{caption}
\usepackage{amsthm}
\theoremstyle{remark}

\ifCLASSINFOpdf
\else
\fi
\begin{document}
 \pagenumbering{gobble} 
\title{Recursive Prediction Error Gradient-Based Algorithms and Framework to Identify PMSM Parameters Online}

\author{Aravinda~Perera,~\IEEEmembership{Student Member,~IEEE,}
        Roy~Nilsen 
        

}


\maketitle

\begin{abstract}
Real-time acquisition of accurate machine parameters is of significance to achieving high performance in electric drives, particularly targeted for mission-critical applications. Unlike the saturation effects, the temperature variations are difficult to predict, thus it is essential to  track temperature-dependent parameters online. In this paper, a unified framework is developed for online parameter identification of rotating electric machines, premised on the Recursive Prediction Error Method (RPEM). Secondly, the prediction gradient ($\mathbf{\Psi}^T$)-based RPEM is adopted for identification of the temperature-sensitive parameters, i.e., the permanent magnet flux linkage ($\Psi_m$) and stator-winding resistance ($R_s$) of the Interior Permanent Magnet Synchronous Machine (IPMSM). Three algorithms, namely, Stochastic Gradient (SGA), Gauss-Newton (GNA), and physically interpretative method (PhyInt) are investigated for the estimation gains computation. A speed-dependent gain-scheduling scheme is used to decouple the inter-dependency of $\Psi_m$ and $R_s$. With the aid of offline simulation methods, the main elements of RPEM such as $\mathbf{\Psi}^T$ are analyzed. The concept validation and the choice of the optimal algorithm is made with the use of System-on-Chip (SoC) based Embedded Real-Time Simulator (ERTS). Subsequently, the selected algorithms are validated with the aid of a 3-kW, IPMSM drive where the control and estimation routines are implemented in the SoC-based industrial embedded control system. The experimental results reveal that $\mathbf{\Psi}^T$-based RPEM, in general, can be a versatile technique in temperature-sensitive parameter adaptation both online and offline.

\end{abstract}


\IEEEpeerreviewmaketitle

\section{Introduction}\label{section1}


\IEEEPARstart{A}{t} the wake of electrification in the operational reliability and safety-critical applications such as surface transport, aerospace, and seabed mining, the dependability of the electrical systems becomes of major significance. Also, the increasing urge to reduce the carbon footprint calls upon more efficient power systems.  IPMSM-equipped electric drives become a frontrunner in this context, owing to some of their inherent features such as superior efficiency and power density, thus ease of cooling, design capability for fault-tolerance, and good control dynamics in a wide torque-speed range \cite{Bostanci2017OpportunitiesStudy}.\par
In realizing a high-performance electric drive, the knowledge of exact machine parameters is essential for multiple reasons\cite{Zhu2021OnlineOverview}, yet the exact parameters are often unknown across the operating range. It is, therefore, useful to identify the machine parameters of the electric drive, thus a variety of online and offline identification methods as reviewed in \cite{Zhu2021OnlineOverview, Rafaq2020ARange} have gained attention in recent years. Out of the electric parameters, i.e. $\Psi_m$, $R_s$ and d- and q- axis inductances $L_d$, $L_q$, the first two are temperature-dependent, and display slow dynamics due to the thermal capacity. $L_d$ and $L_q$ can vary rapidly as they are iron core-saturation dependent, a phenomenon that is dictated by the stator current. Nevertheless, simultaneous identification of more than two unknown parameters is prohibited by the rank-deficiency problem of IPMSM \cite{Vaclavek2013ACAnalysis} unless extra efforts are exerted. Adoption of two time-scale routines for fast- and slow- dynamic parameter-sets \cite{Underwood2010OnlineMachines}
 or High-Frequency Signal Injection (HFSI) \cite{Wang2021AnConditions} or a combination of such methods \cite{Xu2014High-FrequencyEffects} have been employed to circumvent the rank-deficiency challenge. In a practical sense, using an offline method to identify the inductances is adequate, because the stator current which affects the inductances, is a measured quantity in electric drives, thus, the inductances can be calculated in real-time. Conversely, sensor-based temperature monitoring is associated with considerable integration- and reliability- concerns, thus indirect temperature tracking is preferred \cite{Wallscheid2021ThermalChallenges}. It is therefore indispensable to identify $\Psi_m$ and $R_s$ online, thus online identification of these critical parameters is focused in this paper, although can be extended to identify the other.
\subsection{Literature Review}
 The RPEM is a set of parameter identification methods presented by Ljung \cite{Ljung1985TheoryIdentification}, in which it is indicated that several well-known techniques like the Recursive Least Squares (RLS) and the Extended Kalman Filter (EKF) methods can be viewed as its subsets. Among these, RLS, perhaps the most widely adopted, is used in \cite{Underwood2010OnlineMachines} for the identification of all electric parameters of PMSM. It is reported in \cite{Brosch2021Data-DrivenMotors}, the use of RLS to improve the performance of the Model Predictive Controlled PMSM by recursively updating the prediction models. In sensorless drives, the position-estimation accuracy is enhanced using RLS in \cite{Morimoto2006MechanicalIdentification, Inoue2011PerformanceIdentification}. A combination of a signal injection scheme and the RLS method is applied in \cite{Xu2014High-FrequencyEffects} to identify IPMSM parameters of a Direct Torque Control drive. The EKF, another popular member of the RPEM family is discussed for online parameter adaptation \cite{Harnefors1996AdaptivePerformance, Li2021GeneralPMSM} offers decent performance at the cost of increased computational burden. An alternative method under RPEM-family, that exploits the sensitivity of the predicted currents to the model parameters has been discussed in \cite{Ljung1985TheoryIdentification}, in which, this method is termed as prediction gradient ($\mathbf{\Psi}^T$)-based Recursive Prediction Error Method (RPEM). $\mathbf{\Psi}^T$-based RPEM offers more consistent estimations \cite{Soderstrom2001SystemIdentification} and the global convergence is more often guaranteed \cite{Brsting1993EstimationMethod, Ljungquist1993RecursiveModels} compared to EKF-based identification. Additionally, opposing to RLS or EKF methods, the digital implementation of $\mathbf{\Psi}^T$-based RPEM can be less demanding due to the possibility of avoiding the tedious computations like the matrix inversions.
\subsection{Research Gaps and Contribution}
In spite of the merits of $\mathbf{\Psi}^T$-based RPEM, it has not been investigated in the last decades, thus omitted in the recent reviews \cite{Rafaq2020ARange, Zhu2021OnlineOverview}.
Another notable research gap in the RPEM-related literature is the absence of basis and underlying principles behind the choices of estimation gains. This article attempts, firstly, to adopt the $\mathbf{\Psi}^T$-based RPEM for online parameter identification of IPMSM, an investigation that has not been done before, to the authors' best knowledge. Three algorithms, namely SGA, GNA and PhyInt become applicable under this context \cite{Ljung1985TheoryIdentification, Nilsen1989Reduced-OrderMachines}. The SGA-based $\Psi_m$ and $R_s$ identification using the offline simulation tools, presented in \cite{Perera2020AOnlineb, Perera2020AOnline} will be extended with the real-time simulation and experimental validation in this article. Similarly, the offline simulation-based GNA investigation in \cite{Perera2020Gauss-Newton:Online} will be extended using the real-time simulation tools and experimental setup in this article. Additionally, PhyInt is also explored for $\Psi_m$ and $R_s$ indentifcation. Eventually, the performances with different algorithms are compared to draw conclusions for optimal algorithm to compute estimation-gains for $\mathbf{\Psi}^T$-based RPEM. Secondly, to fill the absence of an elaborate procedure to identify estimation-gains in the drives domain, a general approach outlined in \cite{Ljung1985TheoryIdentification} is tailored for electric drives with the aim of formulating a thorough and physically insightful framework for RPEM-based identification. The step-by-step sequence explicitly: 1) Choice of Model-Set, $\mathscr{M}$; 2) Choice of experimental conditions; 3) Choice of criterion function; 4) Choice of search direction; 5) Choice of gain-sequence and initial values. In order to focus the scope to parameter identification, a mechanical position-sensor is assumed to obtain the rotor position although the incorporation of position-sensorless schemes within the same scope is possible as shown in \cite{Perera2020AIdentification}. A Zynq System-on-Chip (SoC) based ERTS and a 3 kW-IPMSM experimental setup is used for simulation and experimental validation.
\subsection{Organization}
The paper is organized in the following manner. In Section II, the IPMSM model and control is briefly outlined. The proposed framework and explicit development of $\mathbf{\Psi}^T$-based RPEM is unfolded in the Section III, where the above mentioned sequence is followed. Section IV explores the use of a rotor-speed dependent gain-scheduler to circumvent the cross-coupling effects between $\Psi_m$ and $R_s$. Subsequently, the validation results and discussions are revealed using the ERTS in the Section V and using the experimental setup in the Section VI, while the concluding remarks are contained in the Section VII.

\section{IPMSM Modeling and Control}\label{sec_2_model}
In this section, the dynamic model of the IPMSM and its Field Oriented Control (FOC) is outlined.
The mathematical model of the electrical part of the machine is in the rotor co-ordinates when given in the per-unit (pu) system: 
\begin{IEEEeqnarray}{rCl}\label{eq:1}
\underline{u}_{s}^{r} &=& r_{s}\cdot \underline{i}_{s}^{r}+ \frac{1}{\omega}_{n}\cdot\frac{\mathrm{d} \underline{\psi}_{s}^{r}}{\mathrm{d} t} +\mathbf{j}\cdot n\cdot \underline{\psi}_{s}^{r} \nonumber\\
\underline{\psi}_{s}^{r} &=& \mathbf{x}_s^{r}\cdot \underline{i}_{s}^{r} + \underline{\psi}_{m}^{r}\IEEEyesnumber\\
\nonumber\\
\underline{i}_{s}^{r} &=& \begin{bmatrix}
i_d & i_q\end{bmatrix}^T,
\quad
\underline{\psi}_{m}^{r} = \begin{bmatrix}
\psi_m & 0\end{bmatrix}^T \nonumber\\
\mathbf{x}_s^r &=& \begin{bmatrix}
x_d & 0\\ 
 0& x_q
\end{bmatrix}, \quad
\mathbf{j}=\begin{bmatrix}
0 &-1 \nonumber\\ 
 1& 0
\end{bmatrix}
\end{IEEEeqnarray} 
Here, $u, i, \psi, x, n, \omega_n$ are voltage, current, flux linkage, inductances, electric speed, and nominal rotational frequency respectively. $\vartheta$ is the electrical angle of the mechanical position $\vartheta_{mech} $ whose relationship with $\vartheta$ is given by  $\vartheta = p \cdot \vartheta_{mech}$ where $p$ is the number of pole pairs. Throughout the article, the superscript and subscript denote the reference frame and the location of the quantity (s-stator, r-rotor, m-magnet) respectively. The notation $\hat{}$ and superscript $^*$ indicate the estimated and the reference -quantities respectively.

\begin{figure}[tb]
\centering
\includegraphics[width=0.48\textwidth]{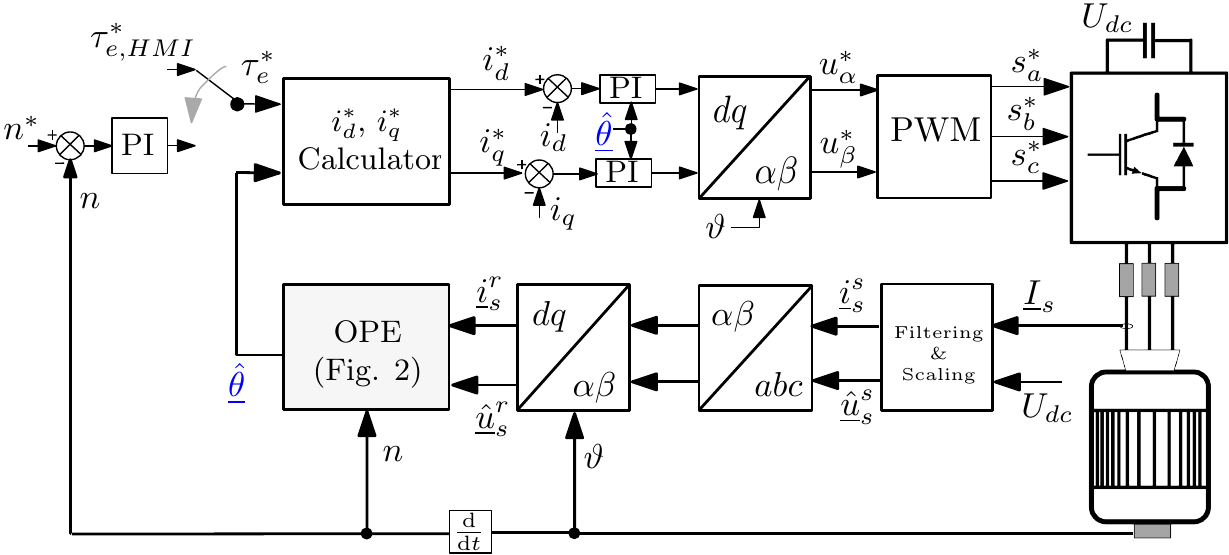}
\caption{{Block Diagram of the Field Oriented Controlled IPMSM Drive enhanced with the Online Parameter Estimator} \label{fig_FOCblockdiagram}}
\end{figure} 

The block diagram of the three-phase, FOC, IPMSM drive enhanced by the Online Parameter Estimator (OPE) is given in Fig. \ref{fig_FOCblockdiagram}. The classical two-level, three-phase Voltage Source Inverter (VSI) is supplied by dc-link capacitors, in which the voltage $U_{dc}$ is measured and used to estimate the stator winding voltages while compensating for the dead-time effects as given in \cite{Weichbold2001InfluenceControl}. $\underline{I}_s$ is measured at the output of the VSI. The OPE estimates the model parameter vector $\underline{\hat{\theta}}$ that is fed into the reference calculator and Proportional-Integral (PI) controllers. Based on the given torque command $\tau_e^*$ either from the Human Machine Interface (HMI) or from the speed controller, $i_d^*, \:i_q^*$ are calculated to fulfill either MTPA using (\ref{eq_refcurrents}) or the field-weakening strategy at high-speed operations.

\begin{IEEEeqnarray}{rCl}\label{eq_refcurrents}
i^*_d &=& \frac{\frac{\hat\psi_m}{3} - \sqrt[3]{(\frac{\hat\psi_m}{3})^3+ \frac{(\hat{x}_q-\hat{x}_d)^2\cdot (\tau^*_e)^2}{3\cdot \hat\psi_m}}} {\hat x_q - \hat x_d} \nonumber\\
 i^*_q &=& \frac{\tau^*_e}{\hat \psi_m - (\hat x_q - \hat x_d)\cdot i^*_d} \IEEEyesnumber
\end{IEEEeqnarray}

\section{Proposed Framework and Development of $\mathbf{\Psi}^T$-based RPEM}\label{section3}
To begin with, RPEM can be generalized as in (\ref{eq_RPEM_intro}).
\begin{IEEEeqnarray}{rCl}\label{eq_RPEM_intro}
     \underline{\hat{\theta}}[k] &=&\begin{bmatrix}
\underline{\hat{\theta}}[k-1] + \mathbf{L}[k,\underline{\hat{\theta}}]\cdot\underline{\epsilon}[k,\underline{\hat{\theta}}]
\end{bmatrix}_{D_{\mathscr{M}}}
\end{IEEEeqnarray}
Here, $\mathbf{L}$ is the gain-matrix, $\underline{\epsilon}$ is the criterion function that we attempt to minimize and eventually nullify, choosing appropriate $\mathbf{L}$. Among the various approaches to compute $\mathbf{L}$, we adopt $\mathbf{\Psi}^T$-based methods. Opposing to the common practice, in this section, we aim to reveal the underlying principles of computing $\mathbf{L}$ by adopting the step-by-step approach \cite{Ljung1985TheoryIdentification} from, for online identification of three-phase IPMSM parameters.
\subsection{Choice of Model-Set, $\mathscr{M}(\underline\theta)$}\label{sec_ModelSet}
\begin{figure}[t]
\centering
\includegraphics[trim=0 180 0 0,clip, width=0.48\textwidth]{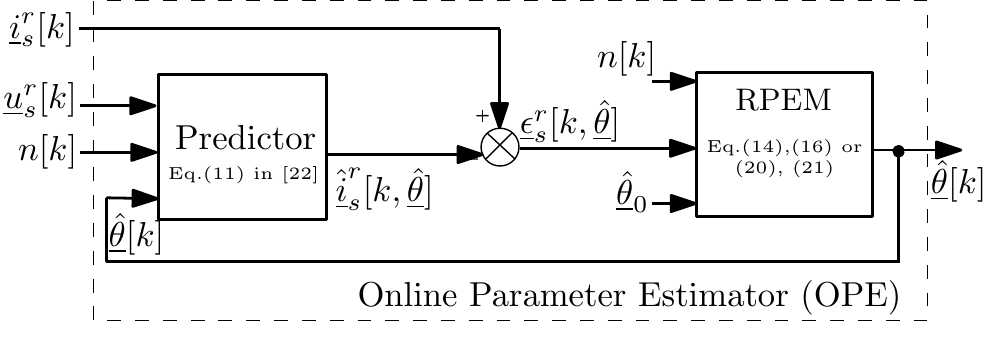}
\caption{{Block diagram of the Open-Loop Online Parameter Estimator}\label{fig_OPEblockdiagram} } 
\end{figure}

The Full-Order Model, $\mathscr{M}_{u\theta}$, given by (\ref{Eq_FullOrderModel}) is chosen under the proposed method because it incorporates the electric parameters of interest. $\mathscr{M}_{u\theta}$ is used to construct a predictor to predict the stator current, $\underline{\hat{i}}^r_s$. The prediction-error, $\underline{\epsilon}^r_s$ is then generated using the measured and the predicted currents, which can be expressed in discrete form as $\underline{\epsilon}^r_s = \underline i^r_s[k] - \hat{\underline{i}}^r_s[k ,\underline{\hat{\theta}}]$. It is assumed that the sole cause for nonzero $\underline{\epsilon}^r_s$ is the difference between the physical and model parameters. The block diagram of the $\mathscr{M}_{u\theta}$-based OPE is given in the Fig. \ref{fig_OPEblockdiagram}. $\underline{\epsilon}^r_s$ is fed forward instead of feedback correction mechanism, unlike in a closed-loop/observer structure. Therefore, this open-loop predictor arrangement enriches $\underline{\epsilon}^r_s$ with parameteric error information, a feature that is attempted to capitalize in computing the prediction gradients under this method. $\underline{\epsilon}^r_s$ is discussed in detail in the Section \ref{Head_CritFunc}.

\begin{IEEEeqnarray}{rCl}\label{Eq_FullOrderModel}
\underline{u}_{s}^{r} &=& \hat{r}_{s}\cdot \underline{i}_{s}^{r}+ \frac{\mathbf{x}_s^r}{\omega}_{n}\cdot\frac{\mathrm{d} \underline{i}_{s}^{r}}{\mathrm{d} t} + \mathbf{j}\cdot n\cdot\mathbf{x}_s^r\cdot \underline{i}_{s}^{r} +\mathbf{j}\cdot n\cdot \hat{\underline{\psi}}_{m}^{r} \nonumber \qquad\\
\underline{i}_s^r &=& \mathbf{T}_{ss}^r(\vartheta)\cdot \underline{i}_s^s, \qquad
\underline{u}_s^r = \mathbf{T}_{ss}^r(\vartheta)\cdot \underline{u}_s^s \IEEEyesnumber
\end{IEEEeqnarray}

\begin{figure}[tb]
\begin{minipage}[t]{0.5\linewidth}
    \subfloat[]{\includegraphics[width=\linewidth]{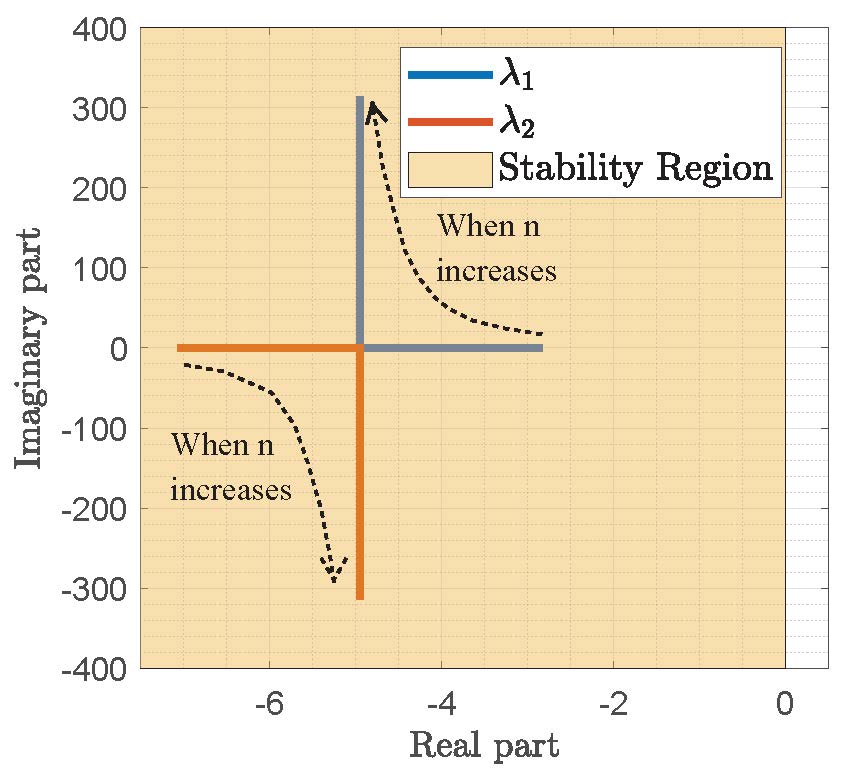}}
\end{minipage}%
    \hfill%
\begin{minipage}[t]{0.5\linewidth}
    \subfloat[]{\includegraphics[width=\linewidth]{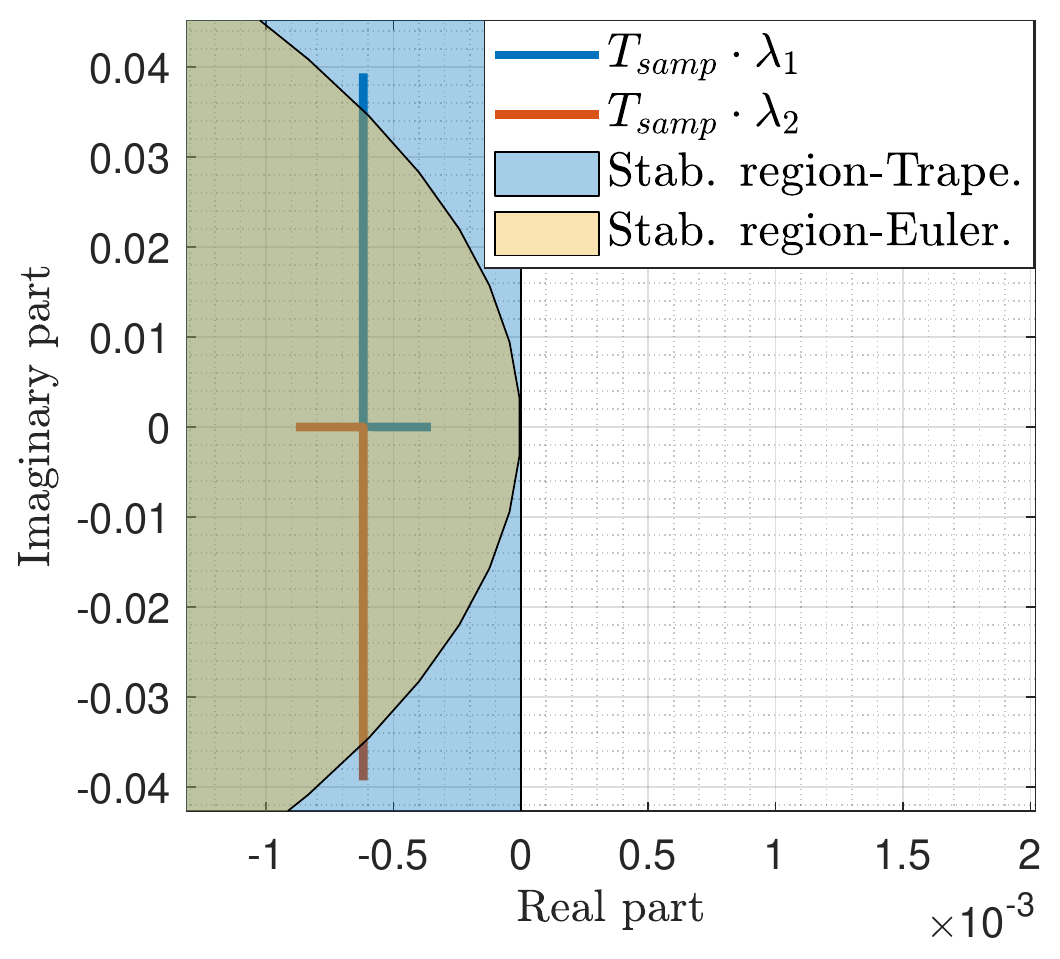}}
\end{minipage}
\caption{Eigenvalue trajectories and stability regions (a) in continuous time domain (b) when discretized using trapezoidal method, in $\lambda-T_{samp}$ plane}
\label{Fig_EigenTraj}
\end{figure}

$\mathscr{M}_{u\theta}$ is a second-order system in which the linearized system matrix \textbf{A} and the eigenvalues, $\lambda_{1,2}$ are given in the (\ref{Eq:SystemMatrix}) and (\ref{Eq:lambda}) respectively, where $\hat{T}_d, \hat{T}_q$, expressed in (\ref{Eq:TdTq}) are $d,q$- axes time-constants. Fig. \ref{Fig_EigenTraj}(a) plots the trajectories of $\lambda_{1,2}$ against the increasing rotor speed from standstill for the IPMSM given in the Table \ref{tab_machineData}. It is evident that the $\mathscr{M}_{u\theta}$ is stable across the full speed range yet, $\underline{\hat{i}}^r_s$ can contain oscillations and their frequency is expected to increase in proportion to the rotor speed \cite{Perera2022InvestigationPress}. Due to this speed-dependency, the numerical method adopted to discretize $\mathscr{M}_{u\theta}$ as well as the integration time-step of the digital controller can influence the stability of the digitally implemented predictor. It is validated in \cite{Perera2022InvestigationPress} that, unlike the explicit Euler method, the trapezoidal rule based numerical method can guarantee the full speed-range stability of $\mathscr{M}_{u\theta}$-based open-loop predictor when implemented in a processor at sampling times $T_{samp}$ corresponding to IGBT-drives. Fig. \ref{Fig_EigenTraj}(b) illustrates, how eigenvalues escape the Euler-based stability region in $\lambda-T_{samp}$-plane, yet are well within that of the trapezoidal rule, when $T_{samp} = 125\,\mu s$.

\begin{IEEEeqnarray}{rCl}\label{Eq:FullOrModAnalysis}
\lambda \cdot I_2 &-& A = \begin{bmatrix}
\lambda + \frac{1}{\hat{T}_d} &  \frac{-n \cdot x_q \cdot \omega_n}{x_d} \nonumber\\ 
 \frac{-n \cdot x_d \cdot \omega_n}{x_q} & \lambda + \frac{1}{\hat{T}_q}\end{bmatrix} \IEEEyesnumber\IEEEyessubnumber\label{Eq:SystemMatrix}\\
 \lambda_{1,2} &=& -\frac{1}{2}\cdot \left ( \frac{1}{\hat{T_d}}+\frac{1}{\hat{T_q}} \right ) \nonumber\\
    &\pm &\sqrt{\left [ \frac{1}{2}\cdot \left ( \frac{1}{\hat{T_d}}+\frac{1}{\hat{T_q}} \right )\right ]^2 - \left [ \left ( \frac{1}{\hat{T_d} \cdot \hat{T_q}} \right )+\left ( \omega_n \cdot n \right )^2 \right ]} \nonumber\\
    \IEEEyessubnumber \label{Eq:lambda}\\
\hat{T}_d &=& \frac{x_d}{\hat{r}_s\cdot \omega_n}, \qquad \hat{T}_q = \frac{x_q}{\hat{r}_s\cdot \omega_n} \IEEEyessubnumber \label{Eq:TdTq}
\end{IEEEeqnarray}

\subsection{Choice of Experimental Conditions}\label{ExpCond}
The choice of experimental conditions imply \textit{when} and \textit{which} data are collected from the process for the identification. The input signals for the OPE model are identified as $\underline{u}^r_s, \underline{i}^r_s, {n}$ as illustrated in the Fig. \ref{fig_OPEblockdiagram}. An online identification method, both at the start-of the drive and during its operation, is chosen as the means to acquire the input signals in order to identify $\Psi_m$ and $R_s$. Hence, $\underline{\hat{\theta}}$ becomes as in (\ref{eq:ParamVectorModified}). In \cite{Ljung1985TheoryIdentification} it is shown that to guarantee global convergence, the $\underline{\hat{\theta}}$ must be bounded by the parameter-space $D_s$, that defines the stable region of the $\mathscr{M}_{u\theta}$-based predictor. To facilitate faster tracking, a narrower parameter-subspace, $D_\mathscr{M}$ can be defined as given in the (\ref{eq:paramsubset}).\par
In addition to the input signals to the OPE, $\vartheta$ needs to be accurately identified, because of the required reference frame transformations in the OPE and FOC in general.

\begin{IEEEeqnarray}{rCl}\label{ParamVector}
      \underline{\hat{\theta}} &=& \begin{bmatrix}
\hat{\psi}_m  & \hat{r}_s
\end{bmatrix}^T, \qquad \underline{\hat{\theta}} \in D_\mathscr{M}, \qquad D_\mathscr{M} \in D_s \qquad \IEEEyesnumber\IEEEyessubnumber\label{eq:ParamVectorModified}\\
 D_\mathscr{M} &=& \begin{Bmatrix}
\hat\psi_{m,min} \leq \hat\psi_m\leq \hat\psi_{m,max} \IEEEyessubnumber\label{eq:paramsubset}\\ 
\hat r_{s,min} \leq \hat r_s\leq \hat r_{s,max}
\end{Bmatrix}
\end{IEEEeqnarray}
\subsection{Choice of Criterion Function, $V_N(\underline{\hat{\theta}})$}\label{Head_CritFunc}
$V_N(\underline{\hat{\theta}})$ and its asymptotic properties are influenced by the choice of $\mathscr{M}(\underline\theta)$. If a Gaussian distribution of the prediction errors is assumed, $V_N(\underline{\hat{\theta}})$ becomes a scalar quadratic criterion \cite{Ljung1985TheoryIdentification} as given in (\ref{Eq:critfunc}) in which $\Lambda$ is the covariance matrix of the prediction error.
\begin{equation}\label{Eq:critfunc}
    V_N(\underline{\hat{\theta}}) = \frac{1}{2} \underline \epsilon_s^{rT}(t,\underline{\hat{\theta}})\cdot\Lambda^{-1}\cdot \underline \epsilon_s^r(t,\underline{\hat{\theta}})
\end{equation}
Assuming that the $\Lambda$ is known and independent of model-parameters, and the prediction error is based on the current measurement, $\Lambda$ is chosen as the Identity Matrix.\par
The sensitivity of the prediction error to all four parametric errors can be evaluated by deriving an expression for the steady-state $\underline{\epsilon}^r_s$ in component form as in (\ref{eq_epsilon_steadystate}).
\begin{IEEEeqnarray}{rCl}\label{eq_epsilon_steadystate}
\epsilon_d &=& -\left( \frac{n^2\cdot \hat{x}_q}{\hat{r}_s^2 + n^2 \cdot \hat{x}_d \cdot \hat{x}_q} \right)\delta\psi_m \nonumber\\ 
&-&\left (\frac{\hat{r}_s}{\hat{r}_s^2 + n^2 \cdot \hat{x}_d \cdot \hat{x}_q} \cdot{i}_d + 
\frac{n\cdot \hat x_q}{\hat{r}_s^2 + n^2 \cdot \hat{x}_d \cdot \hat{x}_q} \cdot {i}_q
 \right) \delta r_s \nonumber\label{eq:ed}\\
 &-&\left(\frac{n^2 \hat{x}_q }{\hat{r}_s^2 + n^2 \cdot \hat{x}_d \cdot \hat{x}_q}\cdot i_d \right)\delta x_d \nonumber\\
&+&\left( \frac{n \cdot \hat{r}_s }{\hat{r}_s^2 + n^2 \cdot \hat{x}_d \cdot \hat{x}_q}\cdot i_q\right)\delta x_q \nonumber\\
\epsilon_q &=&-\left(\frac{n\cdot\hat{r}_s}{\hat{r}_s^2 + n^2 \cdot \hat{x}_d \cdot \hat{x}_q} \right)\delta\psi_m \nonumber\\ 
&-&\left (\frac{\hat{r}_s}{\hat{r}_s^2 + n^2 \cdot \hat{x}_d \cdot \hat{x}_q} \cdot {i}_q - 
\frac{n\cdot \hat x_d}{\hat{r}_s^2 + n^2 \cdot \hat{x}_d \cdot \hat{x}_q} \cdot {i}_d
 \right) \delta r_s \nonumber\\
&-&\left(\frac{n \cdot \hat{r}_s }{\hat{r}_s^2 + n^2 \cdot \hat{x}_d \cdot \hat{x}_q}\cdot i_d \right)\delta x_d \nonumber\\
&-&\left(\frac{n^2 \cdot \hat{x}_d }{\hat{r}_s^2 + n^2 \cdot \hat{x}_d \cdot \hat{x}_q}\cdot i_q \right)\delta x_q \nonumber\\
 \delta\psi_m&=& \psi_m - \hat{\psi}_m, \qquad \delta r_s = r_s - \hat{r}_s \nonumber \\
 \delta x_d&=& x_d - \hat{x}_d, \qquad  \delta x_q= x_q - \hat{x}_q
\IEEEyesnumber\label{eq:deltaParam}
\end{IEEEeqnarray}

\begin{figure}[tb]
\begin{minipage}[t]{0.5\linewidth}
    \subfloat[]{\includegraphics[width=\linewidth]{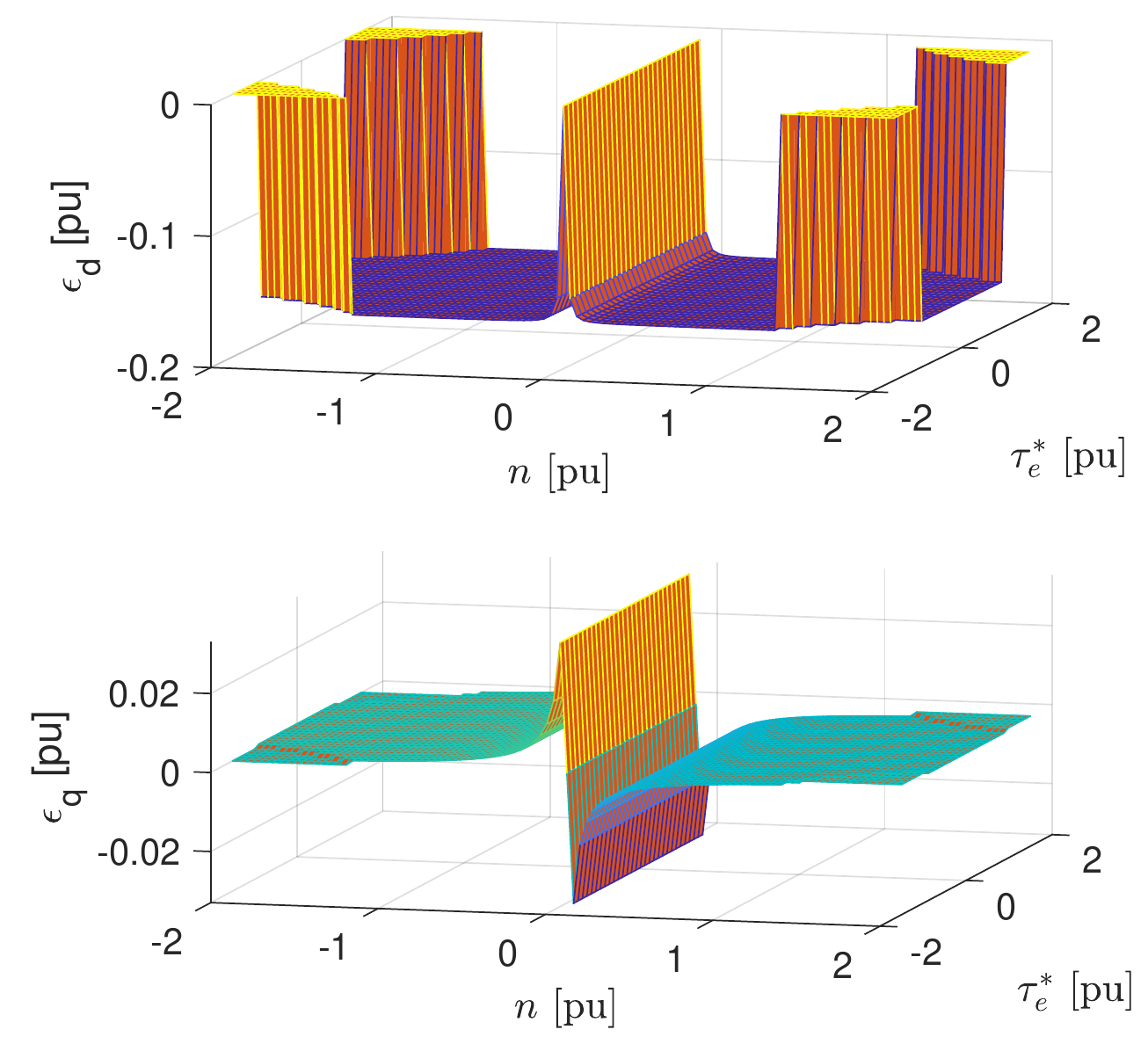}}
\end{minipage}%
    \hfill%
\begin{minipage}[t]{0.5\linewidth}
    \subfloat[]{\includegraphics[width=\linewidth]{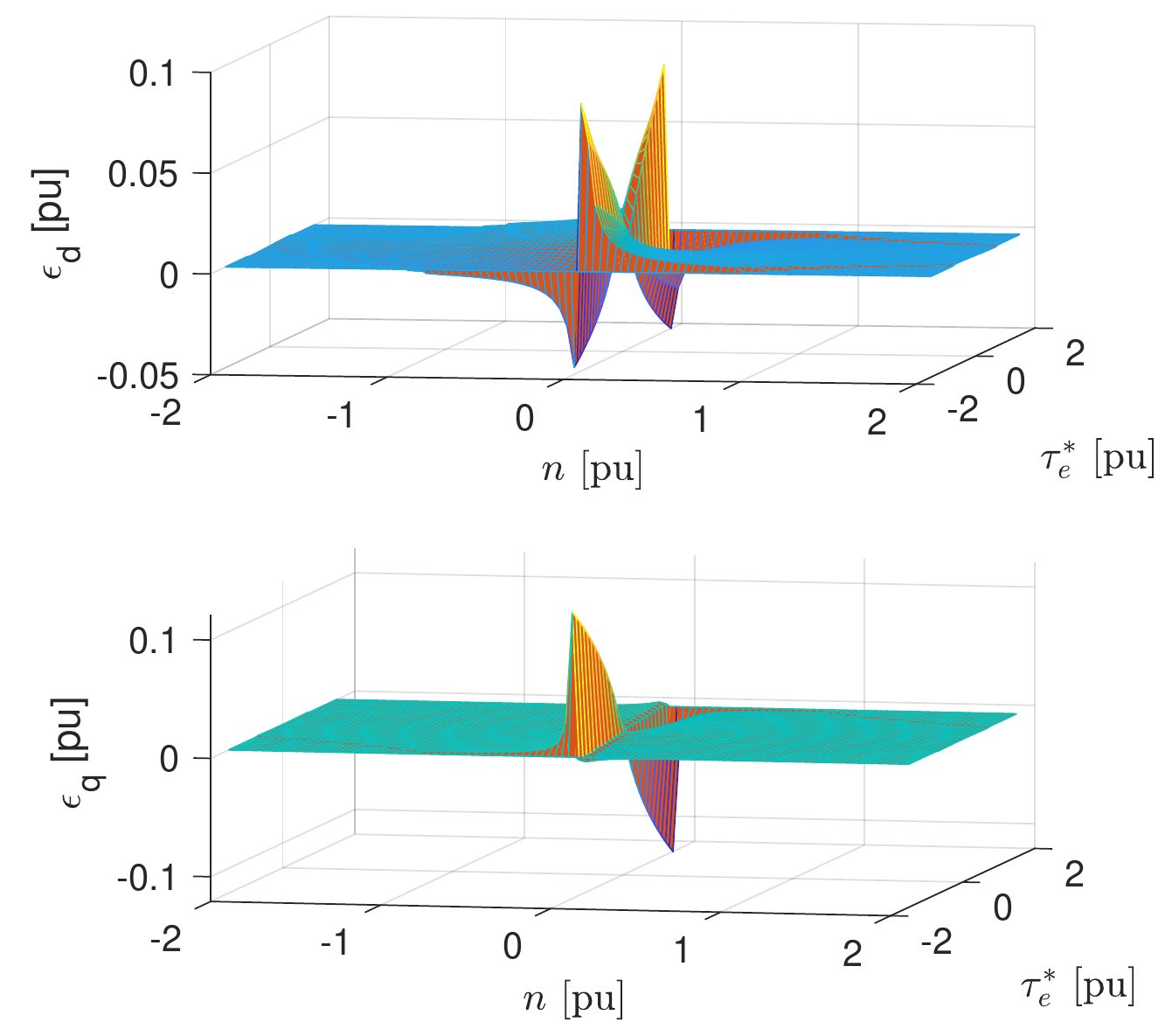}}
\end{minipage}
\caption{Prediction-errors in 4-quadrant speed-torque plane when estimate is 10\% lower than respective physical quantity w.r.t. (a) $\hat{\psi}_m$ (b) $\hat{r}_s$}
\label{Fig_sensitivity}
\end{figure}

To remain within the scope of the article, let us assume the model inductances are in alignment with their physical counterparts, thus $\delta x_d, \delta x_q = 0$ in (\ref{eq_epsilon_steadystate}). Therein, the prediction error sensitivities can be visualized in the 4-quadrant speed-torque plane w.r.t. a 10\% underestimation in $\hat{\psi}_m$ in Fig. \ref{Fig_sensitivity} (a) and a 10\% underestimation in $\hat{r}_s$ in \ref{Fig_sensitivity} (b). In connection to (\ref{eq_epsilon_steadystate}) and Fig. \ref{Fig_sensitivity}, the following observations can be remarked.

 \textit{Remark 1:} When $\delta\psi_m, \: \delta r_s$ becomes zero,  $\epsilon_{d,q}$ also go to zero.
 
 \textit{Remark 2:} When $\delta\psi_m$ is concerned (see Fig. \ref{Fig_sensitivity} (a)) , $\epsilon_d$ is consistently well-condition with $\delta\psi_m$ beyond very low rotor speeds. When $n$ increases, $\epsilon_d \approx \frac{-1}{\hat{x}_d}\cdot\delta\psi_m$. On the contrary, the sensitivity of $\epsilon_q$ to $\delta\psi_m$ across the operating range is weak and inconsistent to make $\epsilon_q$ redundant information for $\psi_m$-identification.
 
 \textit{Remark 3:} When $\delta r_s$ is concerned (see Fig. \ref{Fig_sensitivity} (b)), both $\epsilon_d$ and $\epsilon_q$ become dominant at and around zero-speed to carry rich-conditioned information for $r_s$-identification.
 
 \textit{Remark 4:} When $\delta r_s$ is concerned, $\epsilon_{d,q}$ are also stator current dependent, meaning, even at standstill, $\underline \epsilon^r_s$ carries information to identify $r_s$ if stator current is present.

 \textit{Remark 5:} $\epsilon_{d,q}$ becomes more sensitive to $\delta \psi_m$ and $\delta r_s$ in mutually exclusive speed regions. The dominance of $\delta r_s$-sensitivity is at and around zero speed and this is the very region, the accuracy of $\hat r_s$ becomes critical when the Voltage Model based computations are concerned. 
\subsection{Choice of Search Direction using Prediction Gradient, $\mathbf{\Psi}^T$}\label{Head_PredGradient}
Once $V_N$ is chosen, the correct direction to minimize $V_N$ is discovered using a search direction algorithm. In this article, we focus on algorithms that rely on $\mathbf{\Psi}^T$, which will be developed in this section.\par 
One well-known numerical minimization approach is the use of gradient of the criterion function. It is shown in \cite{Ljung1985TheoryIdentification} that in the pursuit of $\nabla V$, the prediction-error gradient,  $\frac{\mathrm{d} \underline\epsilon^r_s}{\mathrm{d} \underline{\hat\theta}}$ becomes the actual gradient of interest. The prediction-error gradient becomes the negative of the prediction gradient, $\mathbf{\Psi}^T$ as been deduced in (\ref{eq:PredictionGradient})
\begin{IEEEeqnarray}{rCl}\label{eq:PredictionGradient}
\frac{\mathrm{d} \underline\epsilon^r_s[k,\underline{\hat{\theta}}]}{\mathrm{d} \underline{\hat{\theta}}} &=& \frac{\mathrm{d}\underline i^r_s[k,\vartheta] }{\mathrm{d} \underline{\hat{\theta}}} -\frac{\mathrm{d}\underline {\hat i}^r_s[k,\underline{\hat{\theta}}] }{\mathrm{d} \underline{\hat{\theta}}} \nonumber\\
\frac{\mathrm{d} \underline\epsilon^r_s[k,\underline{\hat{\theta}}]}{\mathrm{d} \underline{\hat{\theta}}} &=& -\frac{\mathrm{d}\underline {\hat i}^r_s[k,\underline{\hat{\theta}}] }{\mathrm{d} \underline{\hat{\theta}}} = -\mathbf{\Psi}^T[k,\underline{\hat{\theta}}]  \IEEEyesnumber\label{eq:PG_final}
\end{IEEEeqnarray}
The dynamic forms of the $\mathbf{\Psi}^T$ can be derived by derivation of (\ref{Eq_FullOrderModel}) w.r.t. $\hat{\psi}_m$ and $\hat{r}_s$ as shown in (\ref{eq_PEG_psim_dynamic}) and (\ref{eq_PEG_rs_dynamic}).

 \begin{IEEEeqnarray}{rCll}\label{eq_PEG_psim_dynamic}
 \frac{\mathrm{d} \left(  \frac{\mathrm{d}\hat i_d }{\mathrm{d} \hat \hat \psi_m}\right )}{\mathrm{d} t} &=& 
 \frac{\omega_n}{\hat{x}_d}\Bigg( -\hat r_s\cdot\frac{\mathrm{d}\hat i_d }{\mathrm{d} \hat \psi_m}
  +\hat x_q \cdot n\cdot\frac{\mathrm{d}\hat i_q }{\mathrm{d} \hat \psi_m}  \Bigg) \qquad \IEEEyesnumber\IEEEyessubnumber \\ 
  \frac{\mathrm{d} \left(  \frac{\mathrm{d}\hat i_q }{\mathrm{d} \hat \psi_m}\right )}{\mathrm{d} t} &=& \frac{\omega_n}{\hat{x}_q}\Bigg(-\hat r_s\cdot\frac{\mathrm{d}\hat i_q }{\mathrm{d} \hat \psi_m}
  - \hat x_d \cdot n\cdot\frac{\mathrm{d}\hat i_d }{\mathrm{d} \hat \psi_m} 
 - n \Bigg) \qquad \IEEEyessubnumber
 \end{IEEEeqnarray}

\begin{figure}[tb]
\begin{minipage}[t]{0.5\linewidth}
    \subfloat[]{\includegraphics[width=\linewidth]{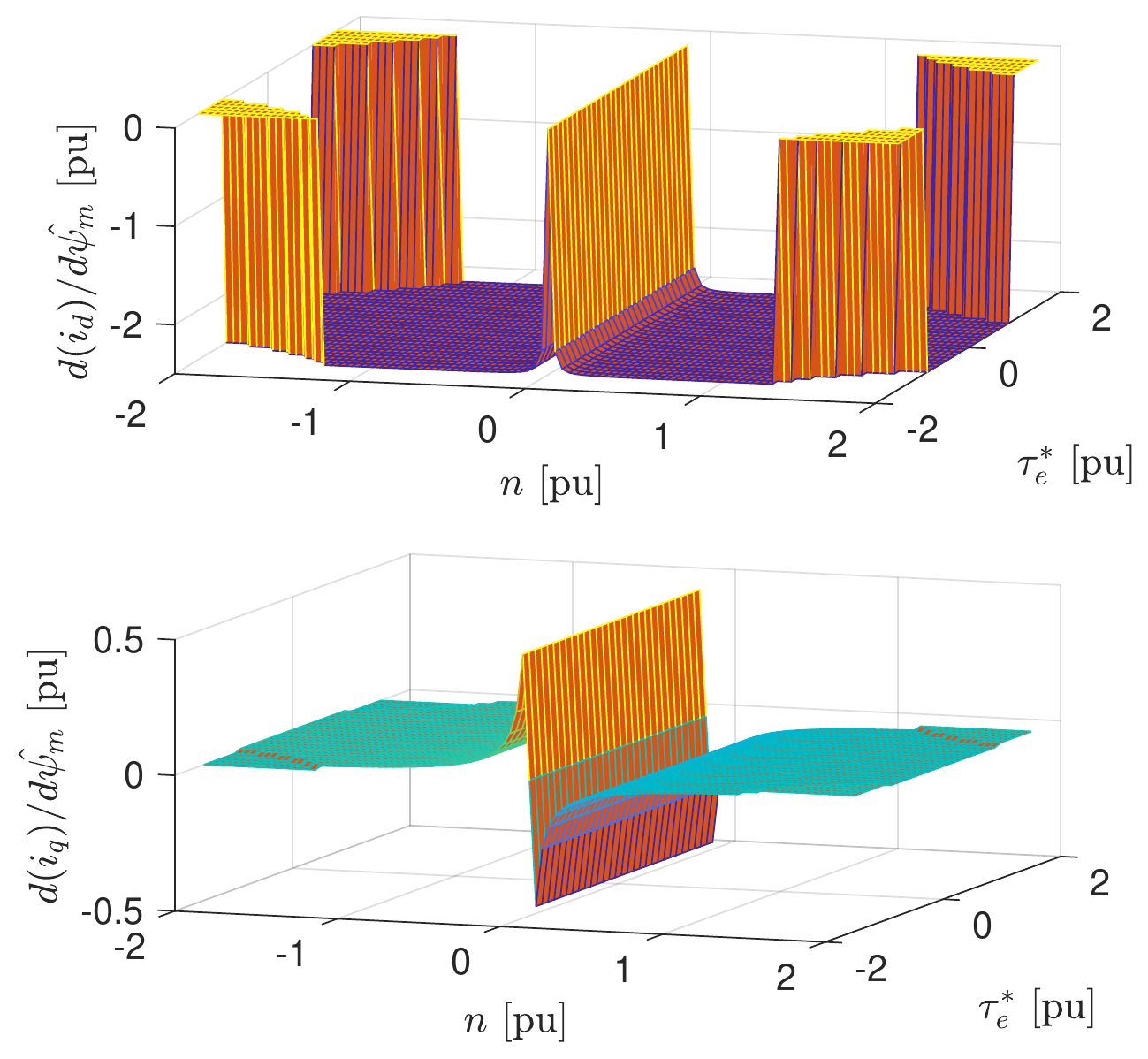}}
\end{minipage}%
    \hfill%
\begin{minipage}[t]{0.5\linewidth}
    \subfloat[]{\includegraphics[width=\linewidth]{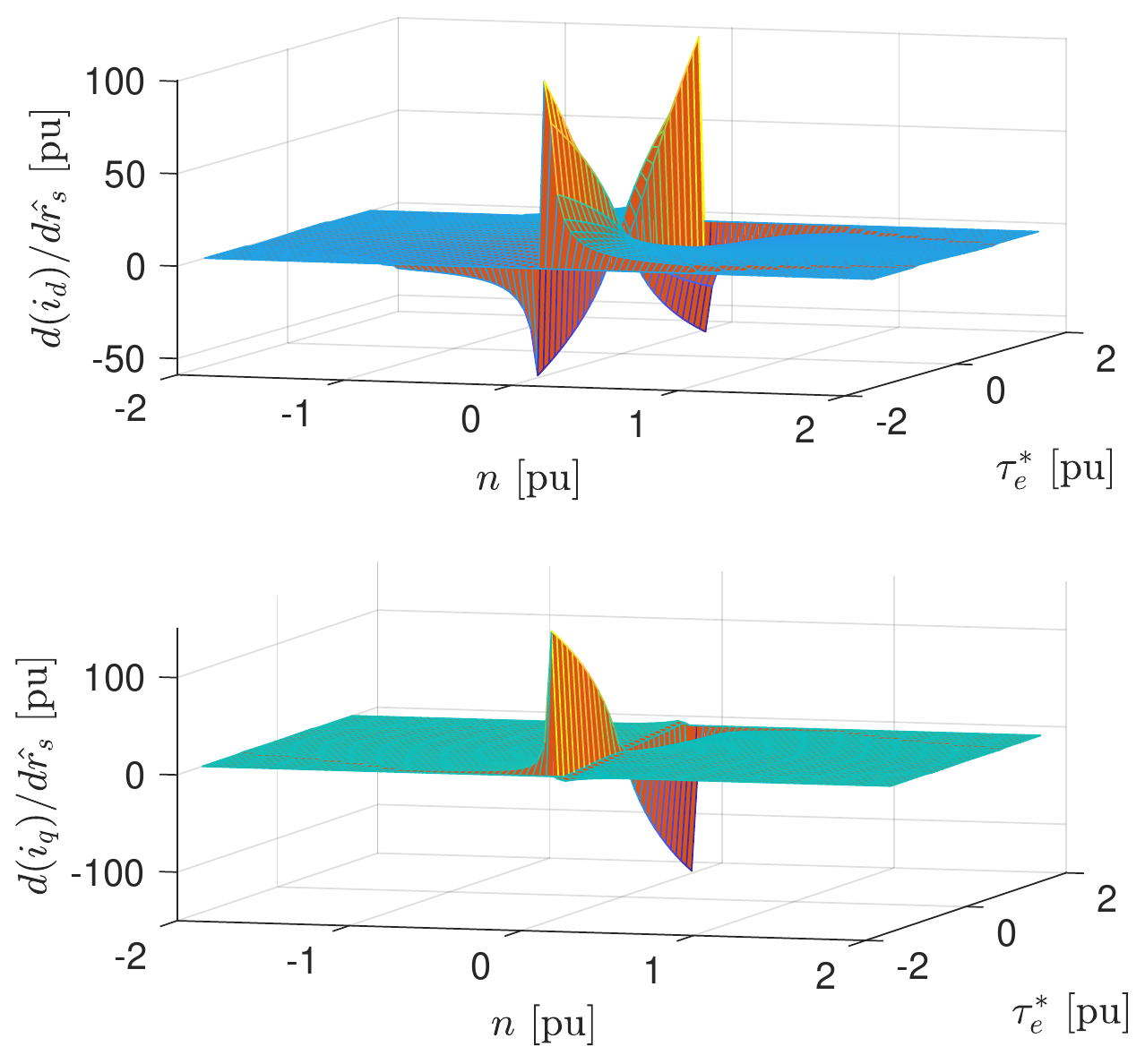}}
\end{minipage}
\caption{{Prediction Gradient in steady-state in 4-quadrant speed-torque plane w.r.t. (a) $\hat\psi_m$ (b) $\hat r_s$} \label{fig_PEG_analytical}}
\end{figure}

 \begin{IEEEeqnarray}{rCl}\label{eq_PEG_rs_dynamic}
 \frac{\mathrm{d} \left(  \frac{\mathrm{d}\hat i_d }{\mathrm{d} \hat r_s}\right )}{\mathrm{d} t} &=& \frac{\omega_n}{\hat{x}_d}\Bigg(-\hat r_s\cdot\frac{\mathrm{d}\hat i_d }{\mathrm{d} \hat r_s} 
 + n \cdot \hat x_q\cdot\frac{\mathrm{d}\hat i_q }{\mathrm{d} \hat r_s} 
 -\hat{i}_d \Bigg)\qquad \IEEEyesnumber\IEEEyessubnumber \\ 
  \frac{\mathrm{d} \left(  \frac{\mathrm{d}\hat i_q }{\mathrm{d} \hat r_s}\right )}{\mathrm{d} t} &=& \frac{\omega_n}{\hat{x}_q} \Bigg( -\hat r_s\cdot\frac{\mathrm{d}\hat i_q }{\mathrm{d} \hat r_s}
  - \hat x_d \cdot n\cdot\frac{\mathrm{d}\hat i_d }{\mathrm{d} \hat r_s} 
 - \hat{i}_q \Bigg) \qquad \IEEEyessubnumber 
 \end{IEEEeqnarray}
 The above dynamic forms of $\mathbf{\Psi}^T$ share the same eigenvalues with $\mathscr{M}_{u\theta}$, thus the concerns regarding the digital implementation discussed in the section \ref{sec_ModelSet} apply to these as well. The corresponding steady-state $\mathbf{\Psi}^T$ forms can be derived by equalizing the the left hand side of the each of the above equations to zero. The final derivations are given in the (\ref{eq:PEG_steadySTate_psim}) and (\ref{eq:PEG_steadySTate_rs}) w.r.t. $\hat \psi_m$ and $\hat r_s$, which can, in fact, be obtained by partially deriving (\ref{eq_epsilon_steadystate}) w.r.t. each parameter estimate. The steady-state $\mathbf{\Psi}^T$-functions are plotted in the Fig. \ref{fig_PEG_analytical}.
\begin{IEEEeqnarray}{lr}
\frac{\mathrm{d} \hat i_d}{\mathrm{d} \hat \psi_m}=-\frac{n^2\cdot{x}_q}{\hat{r}_s+n^2\cdot x_q \cdot x_d} ,\:\frac{\mathrm{d} \hat i_q}{\mathrm{d} \hat \psi_m}=-\frac{n\cdot\hat{r}_s}{\hat{r}_s+n^2\cdot x_q \cdot x_d}\qquad \IEEEyesnumber\label{eq:PEG_steadySTate_psim} \\
\frac{\mathrm{d} \hat i_d}{\mathrm{d} \hat r_s}=-\frac{\hat{r}_s \cdot\hat{i}_d}{\hat{r}_s+n^2\cdot x_q \cdot x_d}  -
\frac{n\cdot x_q\cdot \hat{i}_q}{\hat{r}_s+n^2\cdot x_q \cdot x_d}  \nonumber \\
\frac{\mathrm{d} \hat i_q}{\mathrm{d} \hat r_s}=-\frac{\hat{r}_s\cdot\hat i_q}{\hat{r}_s+n^2\cdot x_q \cdot x_d}  +
\frac{n\cdot x_d \cdot \hat i_d}{\hat{r}_s+n^2\cdot x_q \cdot x_d }  \IEEEyesnumber\label{eq:PEG_steadySTate_rs} 
\end{IEEEeqnarray}
Based on the steady-state functions and corresponding plots, the following remarks can be made.

\textit{Remark 1:} $\mathbf{\Psi}^T$-steady state forms hold the same shapes as their respective $\epsilon$-plots given in the Fig. \ref{Fig_sensitivity}. Their relationship can be explained by (\ref{eq:PredictionGradient}). However, unlike $\epsilon$, $\mathbf{\Psi}^T$ is independent from $\delta \psi_m$ and $\delta r_s$.
 
 \textit{Remark 2:} $\mathbf{\Psi}^T$ w.r.t. $\hat \psi_m$ is excited by $n$. See (\ref{eq:PEG_steadySTate_psim}). In looking at the low derivative due to the inertia, the $n$-excitation can be assumed quite slow, thus, in computation of $\mathbf{L}$ for $\hat\psi_m$ identification, the use of steady-state form of $\mathbf{\Psi}^T$ given in (\ref{eq:PEG_steadySTate_psim}) will be adequate \cite{Perera2020AOnlineb}.
 
 \textit{Remark 3:} $\mathbf{\Psi}^T$ w.r.t. $\hat r_s$ is excited by both $n$ and $\underline {\hat i}^r_s$. See (\ref{eq:PEG_steadySTate_rs}). Dynamic counterparts of $\mathbf{\Psi}^T$ in (\ref{eq_PEG_rs_dynamic}) can offer some sort of a filtering effect in the computed $\mathbf{L}$ owing to the $\hat{T}_d, \hat{T}_q$, while yielding faster adaptation. However, to avoid oscillations in the gain, the steady-state forms can be used instead \cite{Perera2020AOnlineb}.

Now that the $\mathbf{\Psi}^T$-functions are developed, what remains is the choice of $\mathbf{\Psi}^T$-based algorithm. Three algorithms become relevant in this context, namely 1) stochastic gradient 2) Gauss-Newton 3) physically interpretative method, which will be discussed next. The choice of the algorithm is determined by the rate of convergence and the asymptotic accuracy they offer and at which cost of computational tediousness. Note that $\Lambda$ is omitted from the respective expressions due to the same basis associated with (\ref{Eq:critfunc}).
\subsubsection{Stochastic Gradient Algorithm}
This is a rather simple $\mathbf{L}$-computation method given in (\ref{eq:SGA_full}). The algorithm adopts a first-order approximation, i.e. $\mathbf{\Psi}^T$ to identify the search-direction with gains $\gamma[k]$ (later introduced) and $r[k]$, the scalar variant of Hessian Function. Thus, in effect, Stochastic Gradient can be viewed as a modification to the classical gradient descent method. Close inspection of (\ref{eq:SGAHessian}) indicates that $r[k]$ is a first-order filtered version of the traces, (\textit{tr}). Such filtering becomes useful, particularly when dynamic variants of $\mathbf{\Psi}^T$ are applied in computing the \textit{tr}, to prevent undesirable fluctuations in the estimates.
\begin{IEEEeqnarray}{rcl}\label{eq:SGA_full}
\underline{\hat\theta}[k] =\underline{\hat\theta}[k-1] &+& \mathbf{L}[k]\cdot\underline{\epsilon}^r_s[k],\quad \mathbf{L}[k]=\gamma[k]\frac{1}{r[k]}\mathbf{\Psi}[k]\qquad \IEEEyesnumber\IEEEyessubnumber \\
r[k] = r[k-1] &+& \gamma[k] \bigg ( tr \left \{ \mathbf{\Psi}[k]\cdot\mathbf{\Psi}^T[k]\right \}  - r[k-1] \bigg )\qquad \IEEEyessubnumber \label{eq:SGAHessian}
\end{IEEEeqnarray}
 $r[k]$, in steady-state, appears as in (\ref{eq:traceSGAHessian}). 
 \begin{IEEEeqnarray}{rcl}
 tr\left \{ \mathbf{\Psi}[k]\mathbf{\Psi}^T[k]\right \} &=&
\Big(\frac{\mathrm{d} \hat i_d}{\mathrm{d} \hat \psi_m} \Big)^2\!+\!\Big(\frac{\mathrm{d} \hat i_q}{\mathrm{d} \hat \psi_m} \Big)^2+\!\Big(\frac{\mathrm{d} \hat i_d}{\mathrm{d} \hat r_s} \Big)^2+\! \Big(\frac{\mathrm{d} \hat i_q}{\mathrm{d} \hat r_s} \Big)^2\nonumber\\ 
\IEEEyesnumber \label{eq:traceSGAHessian}
 \end{IEEEeqnarray}

\subsubsection{Gauss-Newton Algorithm}
This is, unlike the previous method, a second-order iterative minimization technique, which minimizes the criterion function more efficiently, particularly in the vicinity of the minimum. The simplified algorithm is as in (\ref{eq:GNA_full}).
\begin{IEEEeqnarray}{rCl}\label{eq:GNA_full}
\underline{\hat\theta}[k] &=&\underline{\hat\theta}[k-1] + \mathbf{L}[k]\cdot\underline{\epsilon}^r_s[k],\: \mathbf{L}[k]=\gamma[k]\mathbf{R}^{-1}[k]\mathbf{\Psi}[k] \qquad \IEEEyesnumber\IEEEyessubnumber \\
\mathbf{R}[k] &=& \mathbf{R}[k-1] + \gamma[k]
\Big (\mathbf{\Psi}[k]\cdot\mathbf{\Psi}^T[k] - \mathbf{R}[k-1]\Big) \qquad\qquad
\IEEEyessubnumber
\end{IEEEeqnarray}
Here, the vector form of Hessian, $\mathbf{R}[k]$ is employed. In steady-state, $\mathbf{R}[k] = \mathbf{\Psi}[k]\cdot\mathbf{\Psi}^T[k]$, where the elements of $\mathbf{R}[k]$ become as in (\ref{eq:ExpandedR}).
\begin{IEEEeqnarray}{rcl}\label{eq:ExpandedR}
\mathbf{R} &=&
\begin{bmatrix}
\Psi_{11}^2 + \Psi_{12}^2 & \Psi_{11}\cdot\Psi_{21}+\Psi_{12}\cdot\Psi_{22}\\ 
\Psi_{11}\cdot\Psi_{21}+\Psi_{12}\cdot\Psi_{22}& \Psi_{21}^2 + \Psi_{22}^2
\end{bmatrix}
\end{IEEEeqnarray}
 Owning to the relatively small order of the Hessian, computation of its inverse matrix can be made convenient as in (\ref{eq:InverseR}), by adopting an algebraic manipulation.
\begin{IEEEeqnarray}{rCl}\label{eq:InverseR}
R^{-1}&=&\frac{1}{|\mathbf{R}|}\begin{bmatrix}
R_{22} & -R_{12}\\ 
 -R_{21}& R_{11} 
\end{bmatrix} \nonumber\\
|\mathbf{R}|&=&\Psi_{11}^2\Psi_{22}^2+\Psi_{12}^2\Psi_{21}^2-2\cdot\Psi_{11}\Psi_{12}\Psi_{21}\Psi_{22} \qquad
\end{IEEEeqnarray}

\begin{figure}[t]
\centering
\includegraphics[width=0.48\textwidth]{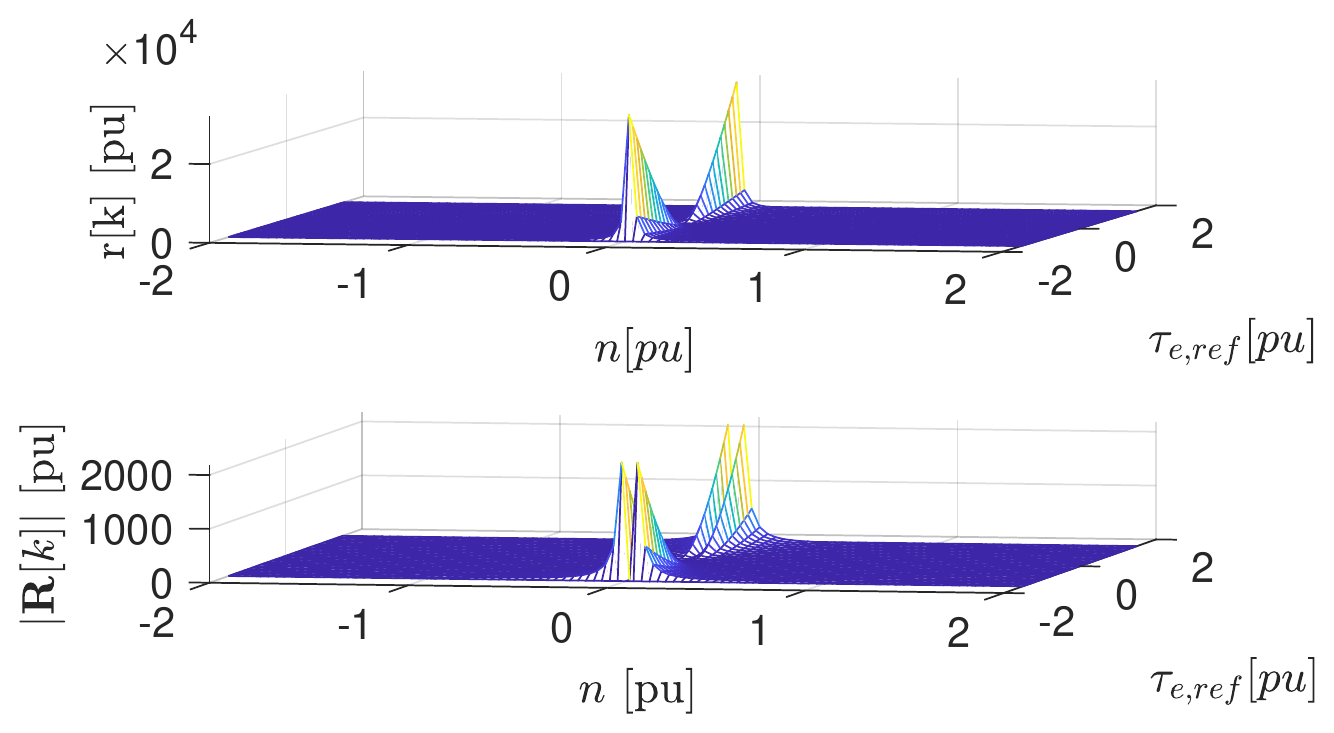}
\caption{{Scalar Hessian, $r$ and the determinant of Matrix Hessian, $|\mathbf{R}|$  in 4-quadrant speed-torque plane}}\label{fig_hesscompstatic}
\end{figure}

In general, Hessian is a function of prediction gradients, and it is independent from $\underline{\epsilon}^r_s$. At zero-speed, $|\mathbf{R}|$ becomes zero so are the elements of $\mathbf{R}$ except $R_{22}$, thus the inverse yields zero-divided-by-zero scenarios in three of its elements. To tackle the challenge with non-existent inverse matrix due to these singularities at zero-speed, a mathematical method called Moore-Penrose pseudoinverse (MPP) is applied to find a pseudoinverse matrix which has most of the properties of $\mathbf{R}^{-1}$\cite{Perera2020Gauss-Newton:Online}. To compare with SGA, the scalar- and the determinant of the matrix- Hessians, which are the denominators of the SGA and GNA, are plotted in the speed-torque plane in the Fig. \ref{fig_hesscompstatic}. In connection to the GNA formulae and Fig. \ref{fig_hesscompstatic}, following remarks can be made.

 \textit{Remark 1:} Both denominators hold similar shapes except at and around zero speed and torque. Despite the similarity in shape, $|\mathbf{R}|$ is several times smaller up to ten times at most at lower speeds, to facilitate faster adaptation with GNA in the lower speed and torque region.
 
 \textit{Remark 2:} At and around zero speed, $r = \psi^2_{21} + \psi^2_{22}$, which is its peak. Conversely, $|\mathbf{R}|$ holds very low values (theoretically zero, but in practice, limited to very low values) to create a cleavage between the peak wedges. In summary, $|\mathbf{R}|$ is expected to offer a larger boost in gain-computation in the lower speed/torque region.
\begin{IEEEeqnarray}{rCl}\label{eq:GNA_GainElements}
\mathbf{L}[k]&=&\frac{\gamma[k]}{|\mathbf{R}|}\begin{bmatrix}
\psi_{11}R_{22}-\psi_{21}R_{12} & \psi_{12}R_{22}-\psi_{22}R_{12}\\  \psi_{21}R_{11}-\psi_{11}R_{12}& \psi_{22}R_{11}-\psi_{12}R_{12}
\end{bmatrix}\quad
 \end{IEEEeqnarray}
From inspection of (\ref{eq:GNA_GainElements}), all the elements in the $\mathbf{L}$ become zero at standstill. This does not influence the $\hat{\psi}_m$- adaptation as the $\underline{\epsilon}^r_s$ anyway does not carry respective information. However, $\underline{\epsilon}^r_s$ does carry information about $\delta r_s$ at zero speed (if is $\underline{i}^r_s \neq 0$) thus forcing $L_{21}, L_{22}$ to null at this point, prevents possible $\hat{r}_s$-adaptation at standstill. This phenomenon indicates an inherent drawback in GNA in comparison to SGA.
\subsubsection{Physically interpretative method}
In this method, the estimation-gains are attempted to obtain by physically interpreting the steady-state behaviour of $\underline{\epsilon}_s^r$ in (\ref{eq_epsilon_steadystate}), s.t. $\mathbf{L}\cdot \epsilon \approx \delta \theta$. We capitalize the physical interpretations in \textit{Remark 2} and \textit{Remark 5} in Section \ref{Head_CritFunc} to identify the estimation-gains. Accordingly, $\hat \psi_m$ estimation-gain becomes:
\begin{IEEEeqnarray}{rCl}\label{eq:PhyInterPRet}
L_{11}[k] &=& -\gamma[k]\cdot \hat{x}_d, \:
\hat\psi_m[k]=\hat\psi_m[k-1]+L_{11}[k]\cdot \epsilon_d[k] \qquad \IEEEyesnumber
 \end{IEEEeqnarray}
 Similarly, the estimation gains for $\hat r_s$-estimation becomes as follows;
  \begin{IEEEeqnarray}{rCl}\label{eq:PhyInterPRet_Rs}
L_{21} &=& \gamma[k] \left(  \frac{\hat r_s^2 + n^2 \cdot \hat x_d \cdot \hat x_q}{-\hat r_s\cdot \hat i_d - n\cdot \hat x_q \cdot \hat i_q } \right)\nonumber\\
L_{22} &=& \gamma[k] \left(  \frac{\hat r_s^2 + n^2 \cdot \hat x_d \cdot \hat x_q}{-\hat r_s\cdot \hat i_q + n\cdot \hat x_d \cdot \hat i_d } \right), \qquad \underline{\hat i}_s[k] \neq 0 \nonumber \\
\hat r_s[k]&=&\hat r_s[k-1]+L_{21}[k]\cdot \epsilon_d[k] +L_{22}[k]\cdot \epsilon_q[k]\IEEEyesnumber
 \end{IEEEeqnarray}
When digital implementation is concerned, SGA, GNA and PhyInt require a minimum value for their denominators ($r$, $|\mathbf{R}|$ or $\underline{\hat i}_s$) at the very low torque/speed region, in order to avoid large $\mathbf{L}$, thus to prevent noise amplification.
\subsection{Choice of Gain Sequence and Initial Values}
Gain-sequence, $\gamma$ can be viewed as a memory-coefficient. Larger $\gamma$ enables faster tracking by 'forgetting' the older $\epsilon$ in preference to the more recent ones however, at the expense of increased noise sensitivity. In the context of tracking slowly varying parameters, it is shown in \cite{Ljung1985TheoryIdentification} that $\gamma[k]$ is often chosen to be a constant, $\gamma_0$, which can be expressed as follows;
\begin{IEEEeqnarray}{rCl}
\underline{\hat{\theta}}[k] &=& \underline{\hat{\theta}}[k-1] + \frac{T_{samp}}{T_0}\cdot\mathbf{\Psi}[k]\cdot\underline\epsilon[k], \quad
\gamma_0 =\frac{T_{samp}}{T_0}
\end{IEEEeqnarray}
Thus, $\gamma_0$ is nothing but an integral time constant, in which, $T_0$ is, in fact, the chosen variable. $T_0$ should be chosen such that the estimated parameters are almost constant over a period of length $T_0$. When temperature-sensitive parameters are concerned, $T_0$ could be in the range of a few seconds such that it still produces a fast enough algorithm to track slow-varying parameters yet not too fast to prevent being sensitive to noise.
Having as much accurate initial values can circumvent a fundamental challenge with the gradient-based minimization algorithms that can be mislead by local minima. By using offline methods and identification runs during the commissioning, initial machine parameters can be identified accurately.
\section{Gain-Scheduling Scheme}\label{Head:Gain-Scheduling}
The inherent coupling of $\Psi_m$ and $R_s$ influences the simultaneous online adaptation. However their impact on each other is not in the same degree \cite{Perera2020AIdentification} and the it was previously revealed that the dominance of sensitivity to each of these parametric errors is in exclusive rotor speed regions, which indicate that an adaptation policy where $\hat r_s$-adaptation happens close to rotor standstill whereas $\hat \psi_m$-adaptation beyond low speeds can be desirable and is encouraged. Thus the respective gains are scheduled as given in (\ref{eq:GainSchedule}), where $x = 1,2$ and $|n_{lim,1}| \geq |n_{lim,2}|$.
\begin{IEEEeqnarray}{rCl}\label{eq:GainSchedule}
L_{1,x} = \left\{\begin{matrix}
L_{1,x},|n| > |n_{lim,1}|\\ 
0, \:otherwise
\end{matrix}\right.,
L_{2,x} = \left\{\begin{matrix}
L_{2,x},|n| < |n_{lim,2}|\\ 
0, \:otherwise
\end{matrix}\right. \qquad
\end{IEEEeqnarray}
\section{Real-Time Simulation Based Validation}\label{sec_simulation}
\begin{figure}[tb]
\centering
\includegraphics[width=0.48\textwidth]{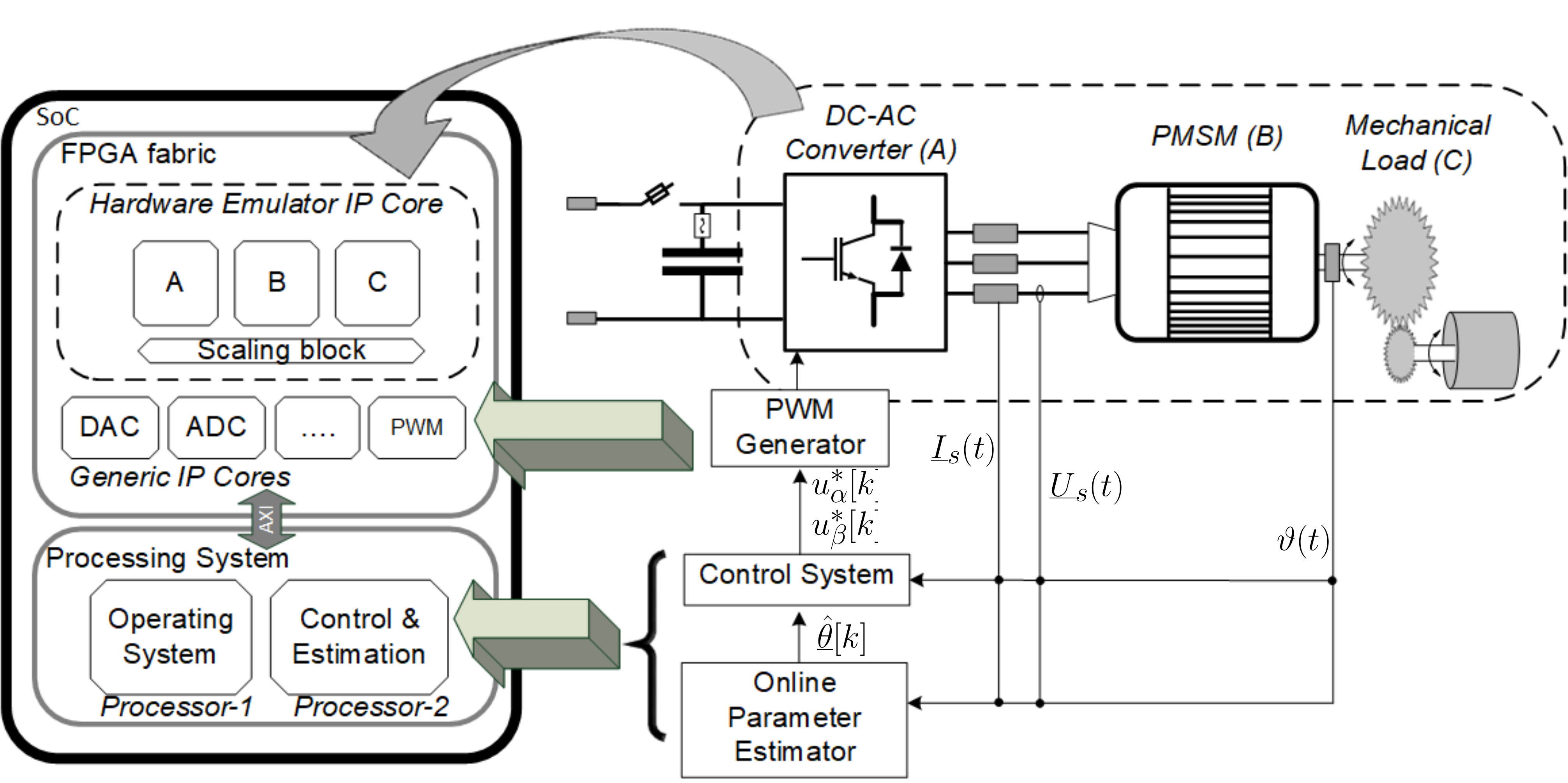}
\caption{{Overview of the Embedded Real-Time Simulator designed for simulation and implementation of three-phase IPMSM Drive}}\label{fig_ERTS}
\end{figure}

In this section, we attempt to make a choice among the three $\mathbf{\Psi}^T$-based algorithms with the aid of a Xilinx Zynq System on Chip-based ERTS. $\hat{\psi}_m$ and $\hat{r}_s$ are identified online when the respective physical values undergo a step-change of -8\% assuming the model inductances are in agreement with their physical counterparts. A step-change in motor parameters allows us to assess the stability and the tracking speed of the proposed method, despite it is unusual for temperature-sensitive parameters. The overview of the ERTS is illustrated in the Fig. \ref{fig_ERTS}. The power hardware components of the drive are programmed in the Field-Programmable Gate Array (FPGA) fabric of the SoC to achieve real-time emulation at a time-step of 1 $\mu s$. The control, state- and parameter- estimation algorithms and likewise relatively slower processes are programmed in the on-chip processor at the PWM double-update time-step of 125 $\mu s$. The validation of this ERTS against the Matlab/Simulink based offline simulation is given in \cite{Perera2021APlatform}. Two-level VSI with asymmetrical modulation and 3$\textsuperscript{rd}$ harmonic injection is used to drive the machine. A speed-dependent gain-scheduler is applied to restrain the $\hat{r}_s$-adaptation between -10 to 10 rpm and $\hat{\psi}_m$-adaptation beyond $|$100$|$ rpm. Table \ref{tab_machineData} tabulates the experimental plant data.
\begin{table}
\caption{Parameters of the Experimental Plant\label{tab_machineData}}
\centering
\begin{tabular}{c c c}
\hline
\hline
Symbol & Parameter & Value\\
\hline
$U_n$ & IPMSM Rated Voltage & 400 V\\
$I_n$ & IPMSM Rated Current & 4.93 A\\
$P_n$ & IPMSM Rated Power & 3 kW\\
$N_n$ & IPMSM Rated Speed & 1000 rpm\\
$T_n$ & IPMSM Rated Torque & 32.6 Nm\\
$p$ & IPMSM Number of pole-pairs & 3\\
$R_s$ & Stator Resistance (offline) & 2.25 $\tcohm$\\
$\Psi_m$ & Permanent magnet flux linkage (offline) & 1.14 Wb\\
$L_d$ & IPMSM d-axis inductance (no-load) & 0.0953 H\\
$L_q$ & IPMSM q-axis inductance (no-load) & 0.206 H\\
$U_{dc}$ & DC bus voltage & 220 V\\
$f_{sw}$ & Power device switching frequency & 4 kHz\\
$T_{samp}$ & Sampling period & 125 $\mu$s \\
\hline
\hline
\end{tabular}
\end{table}

    To avoid the oscillations in the adaptation gains, the steady-state forms of the $\mathbf{\Psi}^T$ is used where applicable \cite{Perera2020AOnlineb}. The respective gain-sequence values for $\hat{\psi}_m$ and $\hat{r}_s$ -adaptation  using SGA, GNA and PhyInt are tabulated in the Table \ref{tab_psimgains}. These values are chosen in order to demonstrate comparable, yet sufficiently rapid tracking performances between the two different algorithms.
\begin{table}[H]
\caption{Gain-sequences for online estimation\label{tab_psimgains}}
\centering
\begin{tabular}{c c c c}
\hline
\hline
Symbol & Parameter & $\gamma_0$[pu] & $\gamma_0$[pu]\\
&& SGA, PhyInt & GNA\\
\hline
&For $\hat{\psi}_m$-estimation&\\
\hline
$\gamma_{0,rk}$ & Gain-sequence for Hessian  & 6.25$\times10^{-4}$ & 6.25$\times10^{-4}$\\
$\gamma_{0,Lk}$ & Gain-sequence for Gain  & 3.25$\times10^{-4}$ & 3.25$\times10^{-4}$\\
\hline
&For $\hat{r}_s$-estimation&\\
\hline
$\gamma_{0,rk}$ & Gain-sequence for Hessian  & 6.25$\times10^{-4}$ & 6.25$\times10^{-5}$\\
$\gamma_{0,Lk}$ & Gain-sequence for Gain  & 6.25$\times10^{-5}$ & 7.5$\times10^{-6}$\\
\hline
\hline
\end{tabular}
\end{table}

\begin{figure}[t]
\centering
\includegraphics[width=0.48\textwidth]{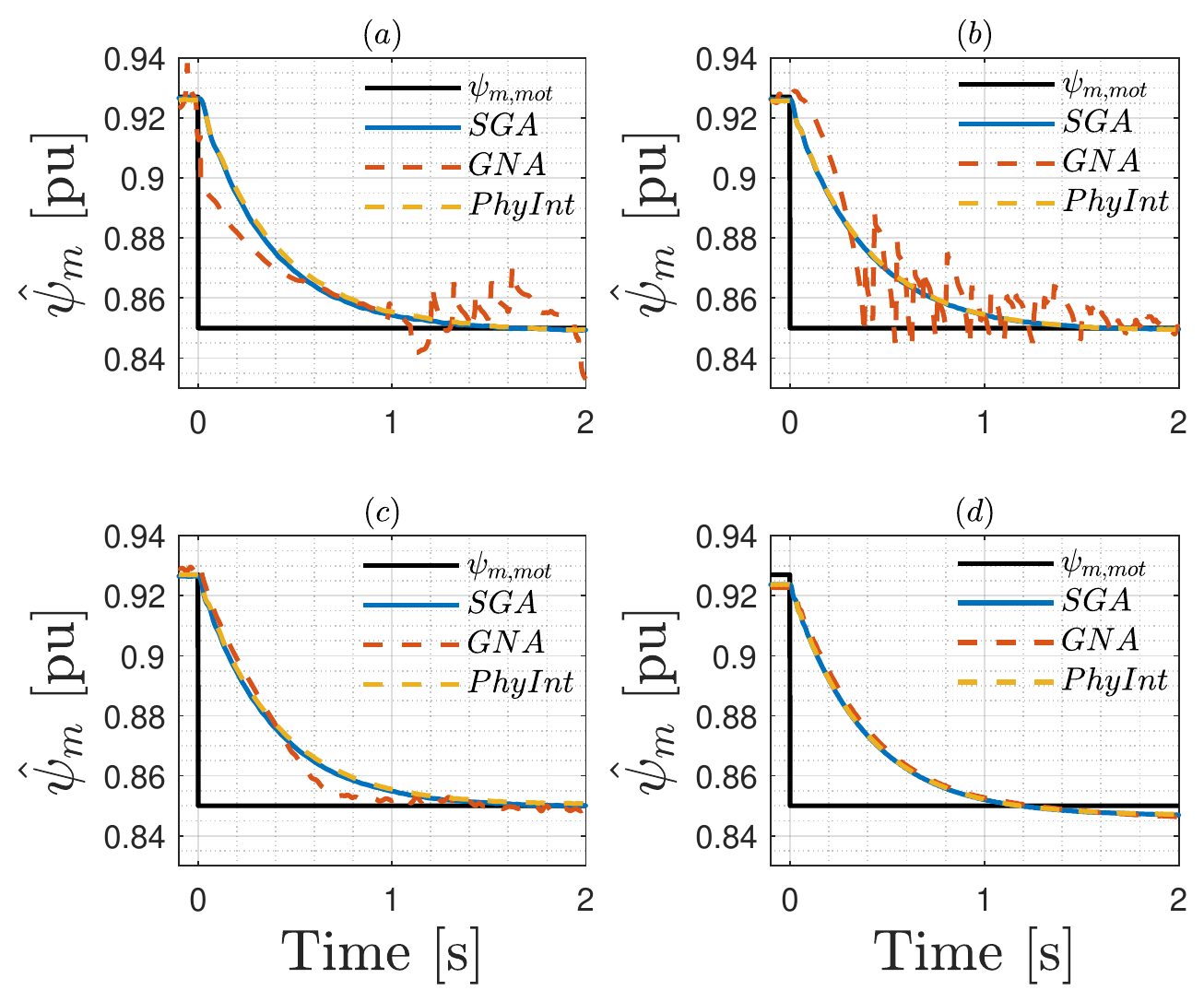}
\caption{{$\hat{\psi}_m$ online adaption with SGA, GNA and PhyInt (a) $n  = -0.2\:pu$ , $\tau_{el} = 0\: pu$ (b) $n  = -0.4\:pu$ , $\tau_{el} = 0.2\: pu$ (c) $n  = 0.4\:pu$ , $\tau_{el} = 0.2\: pu$ (d) $n  = 0.8\:pu$ , $\tau_{el} = 0.4\: pu$}\label{fig_psim_adapt} } 
\end{figure}

\begin{figure}[t]
\centering
\includegraphics[width=0.48\textwidth]{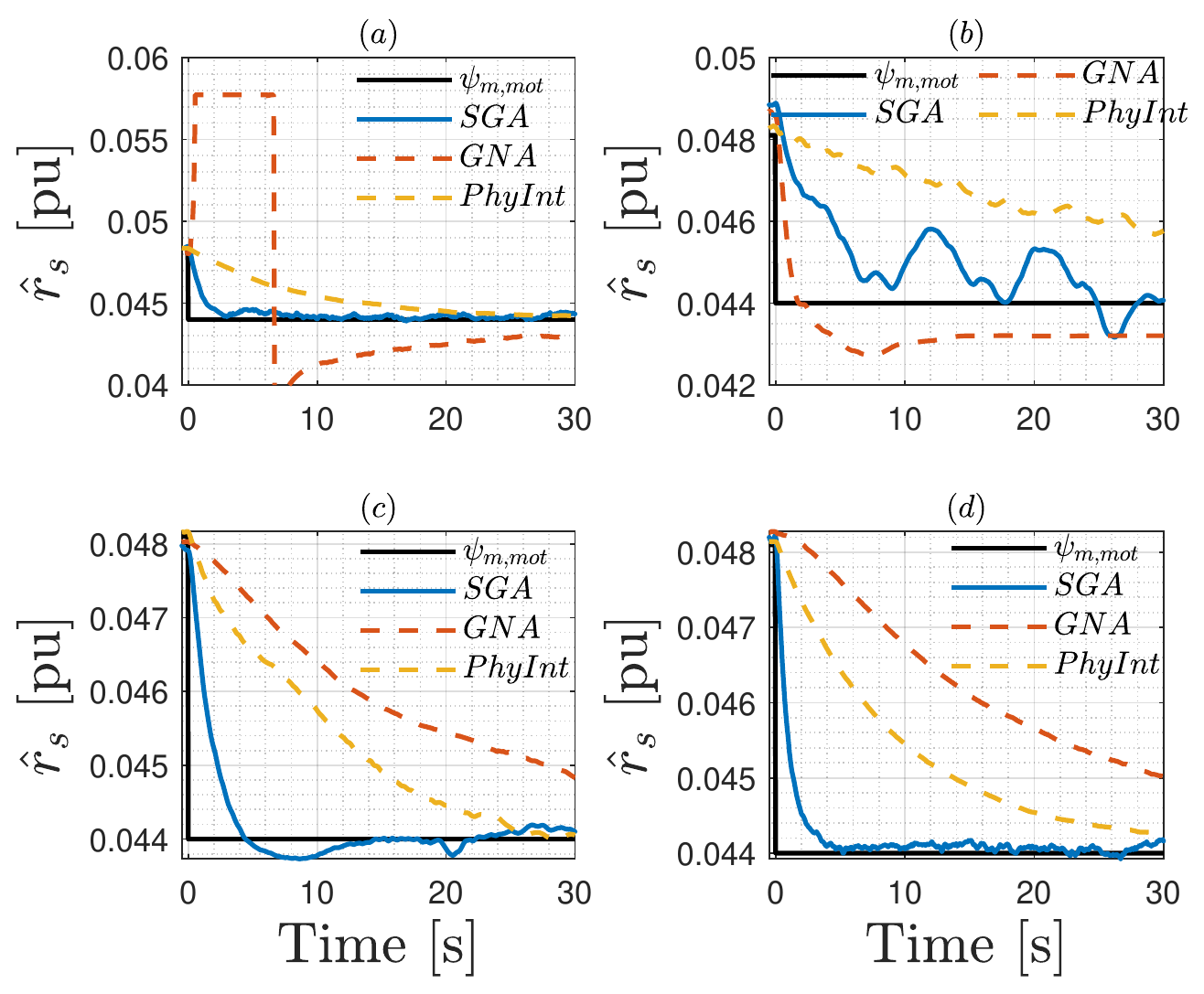}
\caption{{$\hat{r}_s$ online adaption with SGA, GNA and PhyInt (a) $n  = -0.05\:pu$ , $\tau_{el} = 0.2\: pu$ (b) $n  = 0\:pu$ , $\tau_{el} = 0.2\: pu$ (c) $n  = 0\:pu$ , $\tau_{el} = 0.6\: pu$ (d) $n  = 0.05\:pu$ , $\tau_{el} = 0.6\: pu$}\label{fig_rs_adapt} } 
\end{figure}

    Fig. \ref{fig_psim_adapt} contains $\psi_m$ online tracking trajectories overlaid when the three algorithms are adopted at different speeds and loads. Fig. \ref{fig_psim_adapt} (a) and (b) are when the rotor speeds are negative. In case (a) the load-torque $\tau_{el}$ is zero and in (b), $\tau_{el}$ = 0.2 meaning, the machine will be in generating mode, to see nearly no torque in the shaft. Under these conditions, both SGA and PhyInt yield stable and noise-free tracking. GNA, too, succeeds in convergence, yet seem to be overly excited along the way. At low loads, the $\mathbf{R}-$elements in (\ref{eq:InverseR}) become very small which can excessively boost $\mathbf{L}$. This effect is what causes the oscillations in the GNA-trajectories in (a) and (b). At higher loads as in the  Fig. \ref{fig_psim_adapt} (c) and (d), GNA yields smoother adaptation like the SGA and PhyInt.\par
    Similarly, the $R_s$-adaptation related to the three algorithms is presented in the Fig. \ref{fig_rs_adapt}. In this case, to achieve stable tracking with GNA, $\gamma_0$ needed to be made nearly 10 times smaller than that of SGA or PhyInt. This hinders the GNA- tracking speed as it is made evident in all cases. PhyInt, on the other hand, while offering noise-free tracking, the convergence speed is significantly lower in comparison to SGA.\par
    In general, SGA and PhyInt display more stable adaptation consistently. They become the same in steady-state, if $r[k]$ in the SGA is computed using only the respective prediction-gradient instead of the full trace as given in (\ref{eq:traceSGAHessian}). One advantage with SGA over PhyInt is the use of dynamic $r[k]$ (\ref{eq:SGAHessian}) allows initialization and the choice of $\gamma_{rk}$, that can determine the magnitude and length of adaptation-boosting. This facilitates faster and filtered estimations, particularly at start of the routines.
    
\section{Experimental Validation}\label{section6}
\begin{figure}[b]
\centering
\includegraphics[trim=20 150 100 0,clip,width=8 cm, height=4 cm]{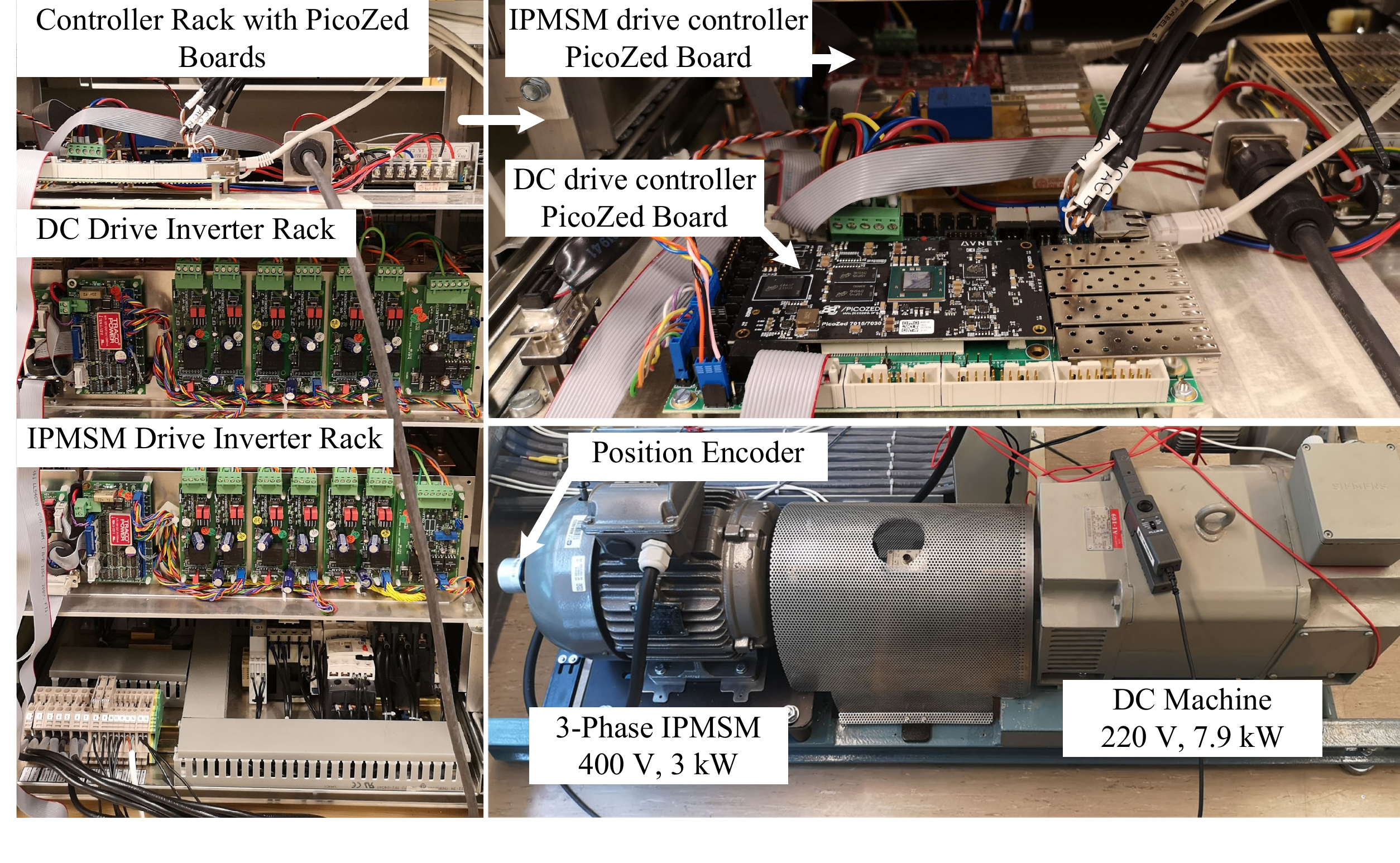}
\caption{{Experimental Setup having IPMSM as the drive machine and DC machine as the load machine mounted on the same shaft}}\label{fig_labsetup}
\end{figure}

Here, the $\mathbf{\Psi}^T$-based RPEM algorithms for parameter identification are attempted to  validate using an experimental setup shown in Fig. \ref{fig_labsetup} of which the data is given in the Table \ref{tab_machineData}. It was evident in the previous section that PhyInt can be viewed as a less flexible variant of SGA, thus, it will be omitted in this experimental validation. The same digital controller that houses the ERTS is used to control the motor drive setup. The $\gamma_0$-values tabulated in Table \ref{tab_psimgains} are applied here.

\subsection{$\Psi_m$-Tracking Validation}
\begin{figure}[t]
\begin{minipage}[t]{0.5\linewidth}
    \subfloat[]{\includegraphics[width=\linewidth]{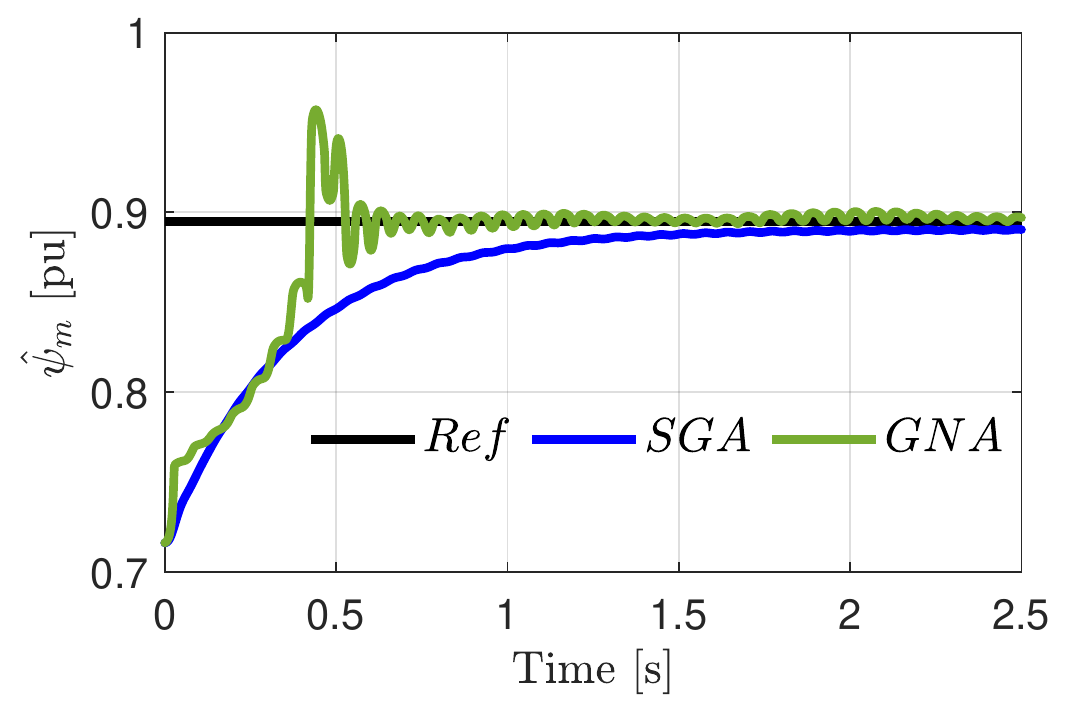}}
\end{minipage}%
    \hfill%
\begin{minipage}[t]{0.5\linewidth}
    \subfloat[]{\includegraphics[width=\linewidth]{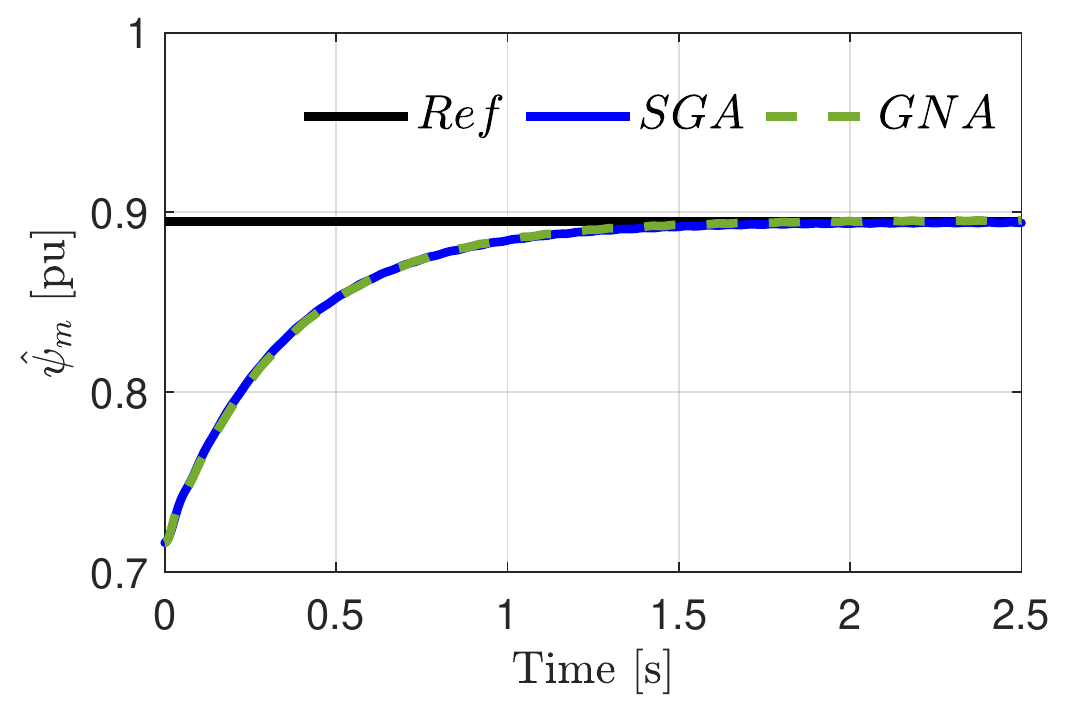}}
\end{minipage}
\\
\begin{minipage}[t]{0.5\linewidth}
    \subfloat[]{\includegraphics[width=\linewidth]{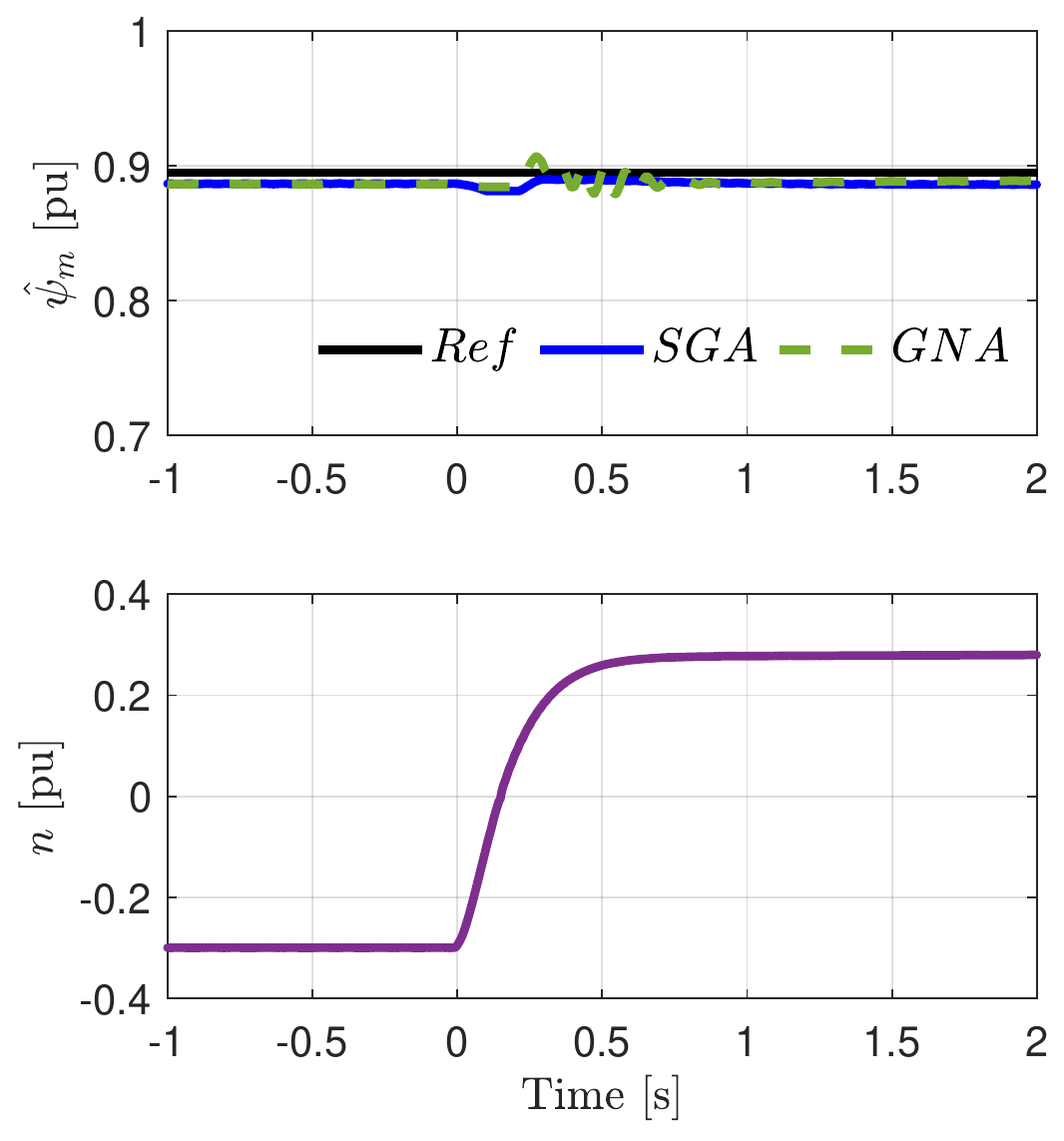}}
\end{minipage}%
    \hfill%
\begin{minipage}[t]{0.5\linewidth}
    \subfloat[]{\includegraphics[width=\linewidth]{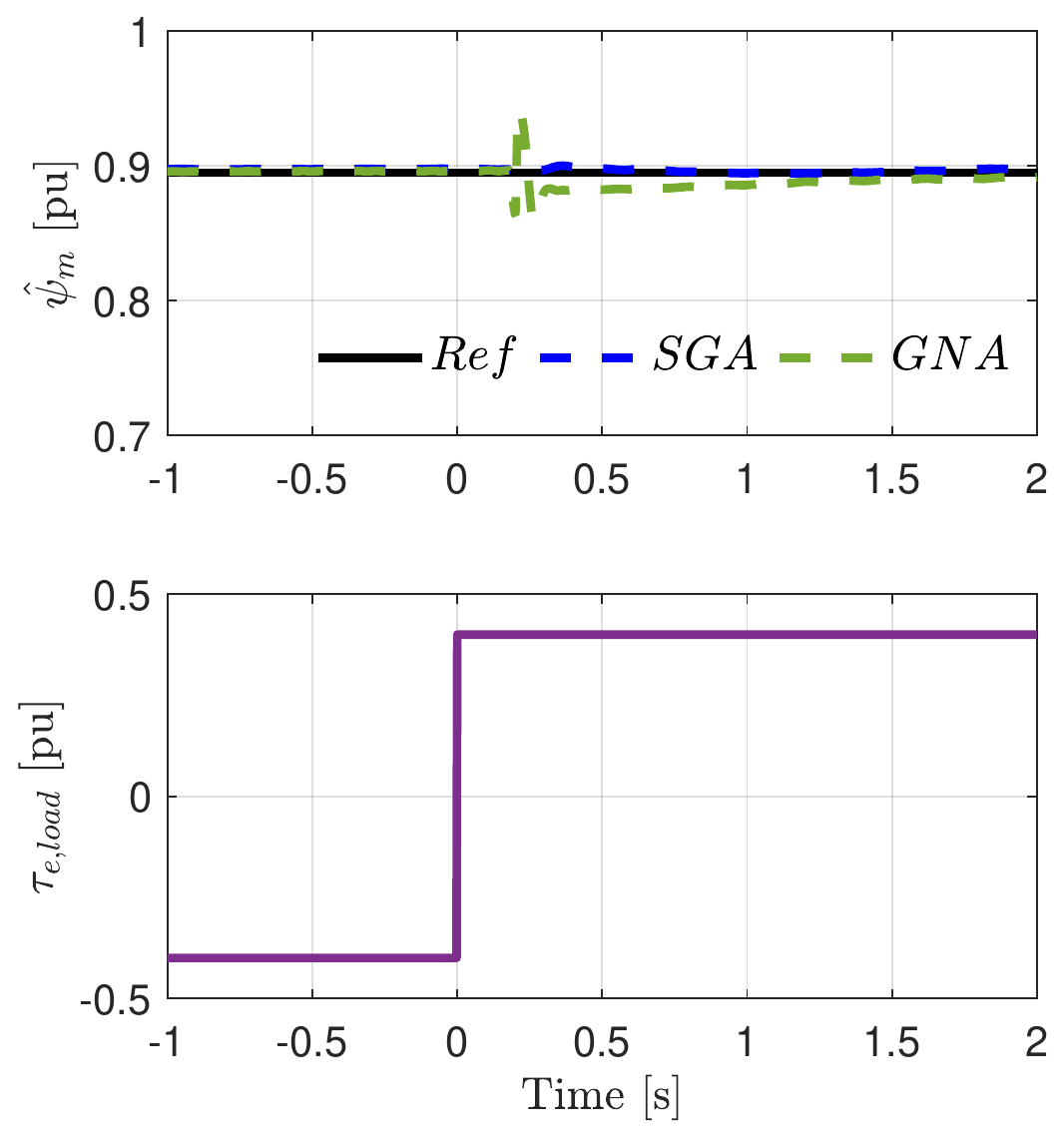}}
\end{minipage}
\caption{Experimental validations of $\hat{\psi}_m$ -online adaptation with SGA and GNA when (a) no-load at 0.3 pu speed
(b) 0.4 pu load-torque at 0.3 pu speed (c) speed reference step-change from -0.3 to 0.3 at 0.4 pu load-torque (d) load step-change from -0.4 to +0.4 pu load-torque at 0.3 pu speed}
\label{Fig_lab_psim}
\end{figure}

During the experiments, it was identified that the dynamic forms of the $\psi_{21}$ and $\psi_{22}$ cause to superimpose the current-sensor noise in the GNA-based tracking trajectories, particularly at the event of no-load. This could have been mitigated by using a 100 times smaller $\gamma_{0,Lk}$ for GNA than what is tabulated in \ref{tab_psimgains}, however at the price of slower convergence. Also, these oscillations disappear as soon as the IPMSM is loaded. Instead, in order to achieve a comparable convergence speed, the steady-state forms of $\psi_{11}$ and $\psi_{12}$ are chosen in both SGA and GNA computations. The performance of the online adaptation of $\hat{\psi}_m$ using these algorithms at various rotor speeds and load torques are plotted in Fig. \ref{Fig_lab_psim}. The reference (Ref) in the plot is the offline identified $\psi_m = 0.895 \, pu$. The no-load adaptation with GNA is slightly quicker than that with the SGA, at the price of a 6\% overshoot, as per Fig. \ref{Fig_lab_psim}(a). When the IPMSM is loaded with 0.4 pu load-torque,  the adaptation between the algorithms is nearly identical as seen in Fig. \ref{Fig_lab_psim}(b). Irrespective of the load, at the given speed, the convergence occurs within 2 seconds which is sufficient for a temperature-induced $\psi_m$-variation.\par
Fig. \ref{Fig_lab_psim}(c) and (d) show how the $\hat{\psi}_m$ behaves upon a step-change in the speed reference and load-torque respectively. In the first case, the speed varies from -0.3 to +0.3 pu speed, during which $\hat{\psi}_m$ remains stable. When a step-change in the load-torque occurs from -0.4 pu to +0.4 pu, i.e. when the sign of the $i_q$ changes, again the $\hat{\psi}_m$ remains stable with the SGA. When GNA is concerned, the $\hat{\psi}_m$ oscillates when the rotor speed is unsettled, yet converges afterward. A summary of the performance is tabulated in the Table \ref{tab_comparison}.

\subsection{$R_s$-Tracking Validation}

\begin{figure}[t]
\begin{minipage}[t]{0.5\linewidth}
    \subfloat[]{\includegraphics[width=\linewidth]{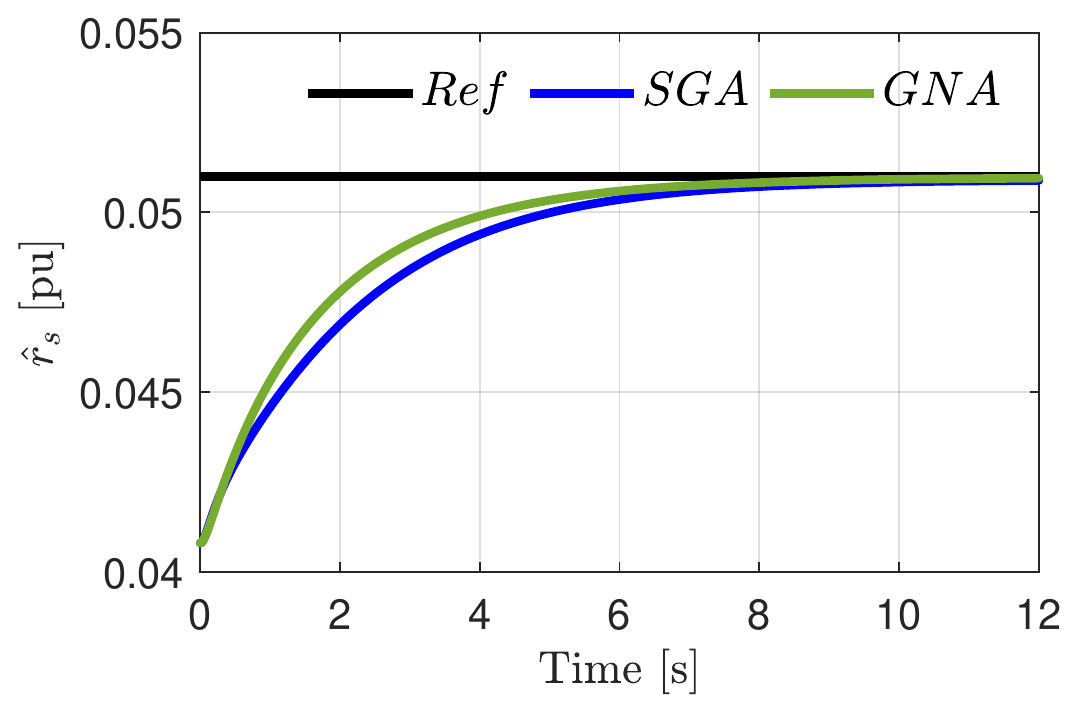}}
\end{minipage}%
    \hfill%
\begin{minipage}[t]{0.5\linewidth}
    \subfloat[]{\includegraphics[width=\linewidth]{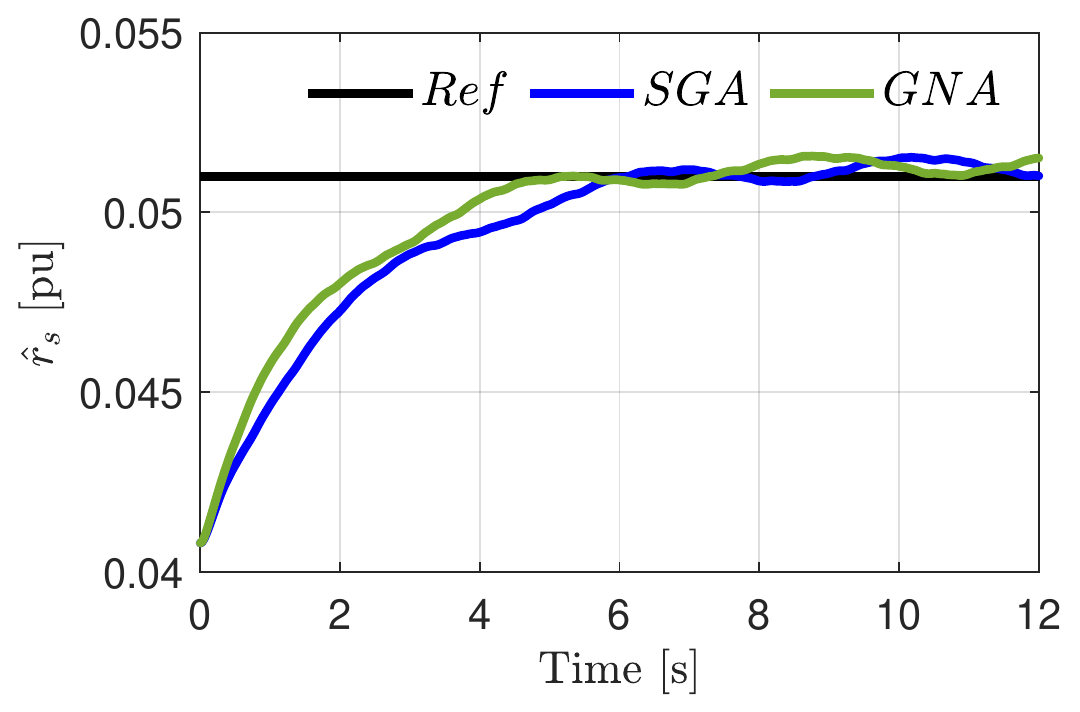}}
\end{minipage}
\\
\begin{minipage}[t]{0.5\linewidth}
    \subfloat[]{\includegraphics[width=\linewidth]{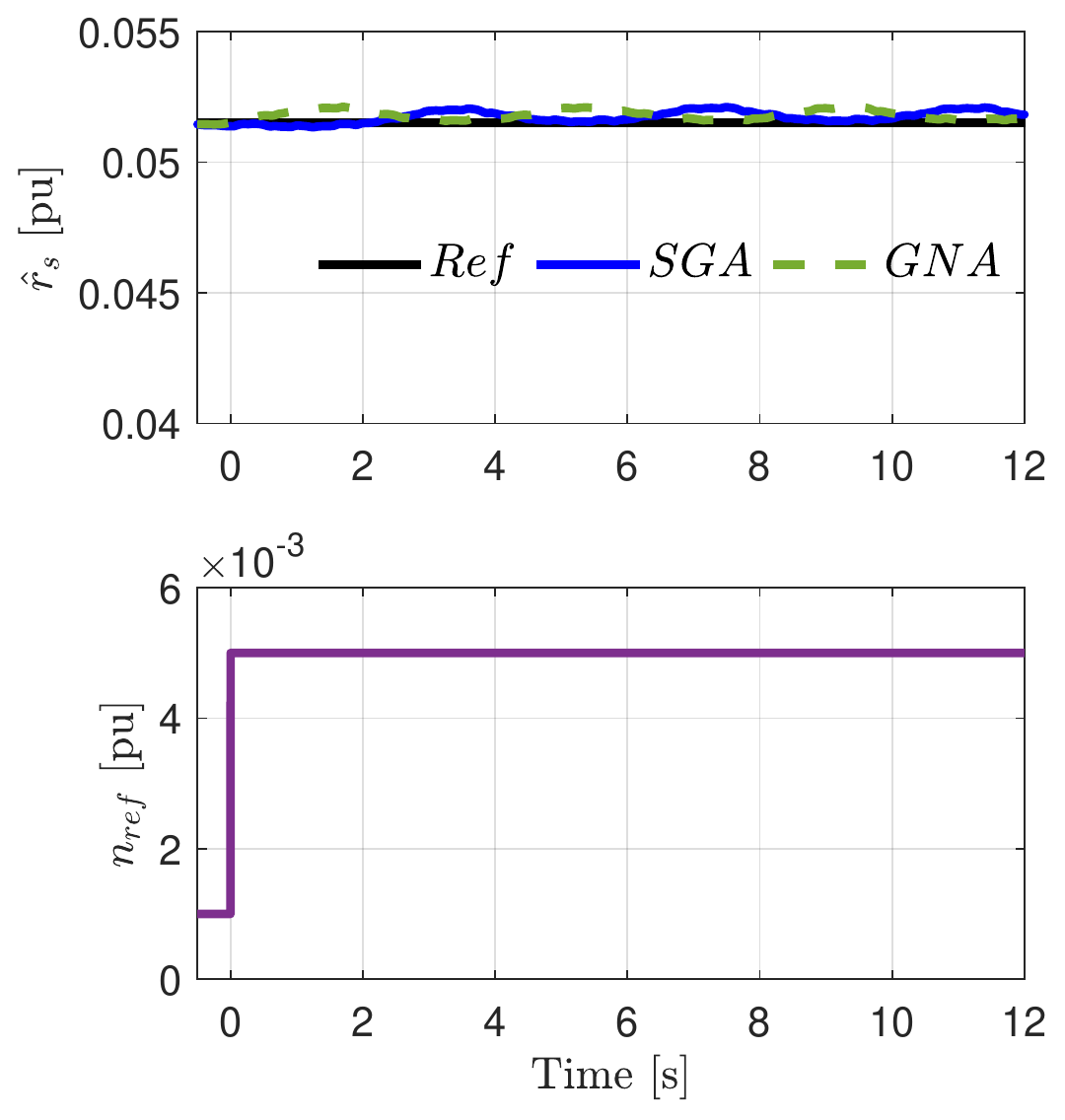}}
\end{minipage}%
    \hfill%
\begin{minipage}[t]{0.5\linewidth}
    \subfloat[]{\includegraphics[width=\linewidth]{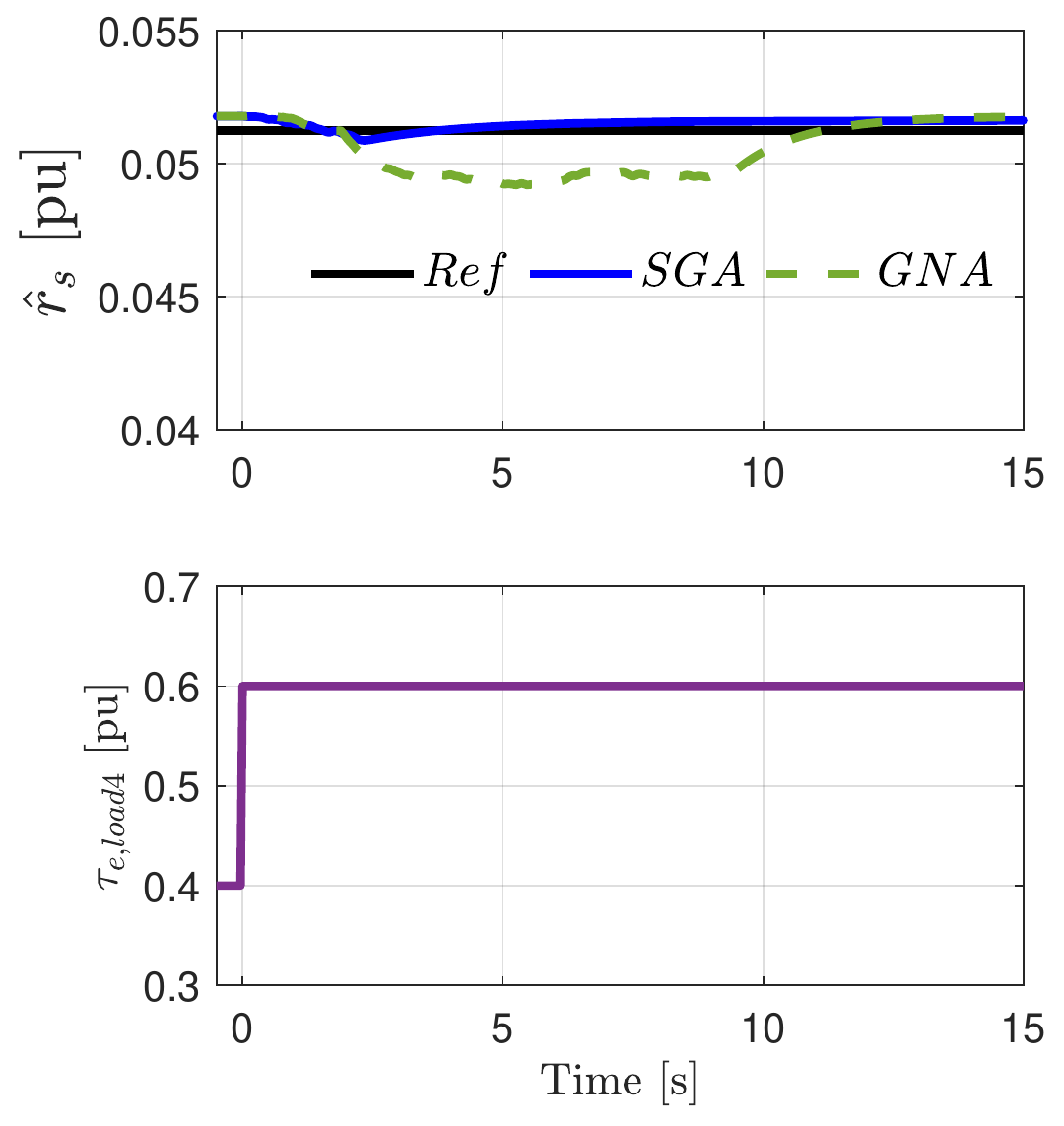}}
\end{minipage}
\caption{Experimental validations of $\hat{r}_s$ -online adaptation with SGA and GNA when (a) 0.4 pu load-torque at standstill
(b) 0.4 pu load-torque at 0.005 pu speed (c) speed reference step-change from 0.001 to 0.005 pu at 0.4 pu load-torque (d) load step-change from 0.4 to 0.6 pu load-torque at standstill}
\label{Fig_Labrs}
\end{figure}

As in the previous case, the steady-state forms of the $\psi_{21}$ and $\psi_{22}$ are incorporated when GNA gains are computed. The respective experimental validations are in Fig. \ref{Fig_Labrs}. The adaption performances at standstill and at 0.005 pu speed are in Fig. \ref{Fig_Labrs} (a) and (b) respectively when the load-torque is 0.4 pu. In both cases, the performance differences between the algorithms are marginal. The convergence performances upon a speed reference and load-torque step-change are plotted in the \ref{Fig_Labrs}(c) and (d) respectively. Despite the steady-state behaviors being indistinguishable, it is seen that the SGA yields more stable tracking during the load (thus the rotor-speed) transient. At low speeds, a speed ripple is evident in the rotor shaft which is superimposed on the estimate-trajectories as seen in the \ref{Fig_Labrs}(b) and (c). A summary of the performance is tabulated in the Table \ref{tab_comparison}. The time taken by the SoC to process SGA and GNA routines is nearly the same ($\sim$20 $\,\mu$s), thus the computational burden is not a matter of concern.

\begin{table}
\caption{Performance comparison summary between SGA and GNA \label{tab_comparison}}
\centering
\begin{tabular}{c c c}
\hline
\hline
Case & SGA & GNA\\
\hline
W.r.t. $\hat{\psi}_m$-estimation\\
\hline
Convergence speed ($\tau_l = 0, n = 0.3 \,pu$)    & $\sim$2 s & $\sim$0.5 s   \\
Convergence speed ($\tau_l = 0.4, n = 0.3 \,pu$)  &  $\sim$1.5 s & $\sim$1.5 s   \\
Steady-state error ($\tau_l = 0, n = 0.3 \,pu$)    & (-)0.5\% & 0.5\%  \\
Steady-state error ($\tau_l = 0.4, n = 0.3 \,pu$)    & $\sim$0\% & $\sim$0\%\\
\hline
W.r.t. $\hat{r}_s$-estimation\\
\hline
Convergence time ($\tau_l = 0.4, n = 0 \,pu$)    & 8 s & 8 s   \\
Convergence time ($\tau_l = 0.4, n = 0.005 \,pu$)  & 6 s & 4 s   \\
Steady-state error ($\tau_l = 0.4, n = 0 \,pu$)    & 0 & 0  \\
Steady-state error ($\tau_l = 0.4, n = 0.005 \,pu$)    & $\sim$0 & $\sim$0\\
\hline
\hline
\end{tabular}
\end{table}

\section{Conclusion}\label{section7}
This article proposed a prediction-gradients-assisted RPEM-based framework and three algorithms to identify parameters of electric machines, and the methods are demonstrated and validated using an IPMSM by identifying temperature-sensitive parameters online. The predictor is arranged in an open-loop thus the prediction error is enriched with parametric errors, a feature that is exploited by deriving prediction gradients, that becomes the main element in the estimation gains in this context. With the aid of real-time simulation, a performance comparison of the three algorithms is executed across the operating range. Experimental results show that both the SGA and GNA offer reasonable tracking performance. Despite the latter can offer faster tracking in principle, it becomes overly excited at lower torque/speed region and inherently prevents $R_s$-tracking at zero-speed, unlike the other two methods. Moreover, very fast adaptation has little use when the large thermal time constants are concerned. Due to the attributes of the dynamic hessian, SGA offers controllable tracking speeds at the start, unlike the PhyInt. Given the stable, flexible, consistent performance and the simplicity in the implementation, RPEM with SGA can be a practical solution for temperature-sensitive parameter estimation of electrical machines.



\ifCLASSOPTIONcaptionsoff
  \newpage
\fi
\bibliographystyle{ieeetr}
\bibliography{references,references2} 
\end{document}